\documentclass{jfm}
\usepackage{graphicx}
\usepackage{epstopdf, epsfig}
\usepackage{natbib}
\usepackage{amsmath} 
\usepackage{amssymb}
\newcommand{\beq}{\begin{equation}}
\newcommand{\eeq}{\end{equation}}

\shorttitle{Helically Decomposed Turbulence}
\shortauthor{A. Alexakis}

\title{Helically Decomposed Turbulence }

\author{Alexandros Alexakis\aff{1}
  \corresp{\email{alexakis@lps.ens.fr}} }

\affiliation{
\aff{1}  Laboratoire de Physique Statistique, \'Ecole Normale Sup\'erieure, PSL Research University; 
         Universit\'e Paris Diderot Sorbonne Paris-Cit\'e; Sorbonne Universit\'es UPMC Univ Paris 06; 
         CNRS; 24 rue Lhomond, 75005 Paris, France }
\begin{document}

\maketitle

\begin{abstract}

A decomposition of the energy and helicity fluxes in a turbulent hydrodynamic flow  is proposed.
The decomposition is based on the projection of the flow to a helical basis that allows 
to investigate separately the role of interactions among modes of different helicity.
The proposed formalism is then applied in large scale numerical simulations of a non-helical and a helical flow,
where the decomposed fluxes are explicitly calculated.
It is shown that the total energy flux can be split in to three fluxes that independently remain constant
in the inertial range. One of these fluxes that corresponds to the interactions of fields with the same helicity
is negative implying the presence of an inverse cascade that is `hidden' inside the forward cascade.
Similar to the energy flux the helicity flux is also shown that it can be decomposed to two
fluxes that remain constant in the inertial range. 
Implications of these results as well possible new directions for investigations are discussed.

\end{abstract}


\section{ Introduction }\label{sec:intro}
%


Hydrodynamic turbulence refers to the state of flow in which eddies self stretch one an other 
to generate a continuous spectrum of excited scales 
from scales of the domain size to scales small enough so that eddies are dissipated by the viscous forces \citep{frisch1995turbulence}.
In its simplest form turbulence is described by the incompressible Navier-Stokes equation that is
going to be considered in this work and is given by: 
\begin{equation}
\partial_t {\bf u}  = \mathbb{P}  \left[ {\bf u \times w} \right] + \nu \Delta {\bf u}   + {\bf F}.
\label{NS}
\end{equation}
where ${\bf u}$ is the three dimensional incompressible ($\nabla \cdot {\bf u}=0$) velocity field and  ${\bf w}$ the vorticity ${\bf w=\nabla \times u}$. 
The domain considered is a triple periodic box of size $2\pi L$.
Energy is injected in the system by the forcing function ${\bf F}$ that acts at some particular length-scale $\ell_f$. 
Dissipation occurs by the viscous forces $\nu \Delta {\bf u}$ where $\nu$ is the  viscosity  coefficient.
$\mathbb{P}$ is the projection operator to incompressible flows that in the periodic domain, that is considered here, can be written as
\beq \mathbb{P}[{\bf u}]\equiv -\nabla \times \nabla \times \Delta^{-1} {\bf u}={\bf u}-\nabla \Delta^{-1}(\nabla\cdot {\bf u})={\bf u}-\nabla P. \eeq 
For a given forcing function $\bf F$ this system has one non-dimensional control parameter 
that is commonly  taken to be the the Reynolds number $\Rey$, 
defined as $\Rey \equiv U\ell_f/\nu$  with $U$ the velocity r.m.s. value.
Turbulence is realized for large values of $\Rey\gg1$ where viscosity becomes effective only at the smallest scales $\ell_\nu \propto \ell_f \Rey^{-3/4} \ll\ell_f$.

The non-linearity of the Navier-Stokes equation conserves energy $E \equiv \frac{1}{2}\left\langle {\bf u \cdot u } \right\rangle$ (where the angular brackets denote space average). 
At steady state the energy injected in the forcing scale $\ell_f\equiv k_f^{-1}$
cascades down to the smallest scales where it is dissipated by the viscous forces. The balance of energy injection to dissipation leads to the relation
\beq
\left\langle{\bf F \cdot u } \right\rangle_{_T}=
\nu \left\langle {\bf w \cdot w}  \right\rangle_{_T}\equiv \epsilon_{_E}
\eeq
where the angular brackets $\left\langle \cdot \right\rangle_{_T}$  denote space and time average. The right hand side expresses the energy injection rate and the left hand side expresses the energy dissipation rate $\epsilon_{_E}$.

The second quadratic invariant conserved by the nonlinearity of the Navier-Stokes equation is the Helicity $H \equiv \frac{1}{2}\left\langle {\bf u \cdot w } \right\rangle$.
Helicity is a topological quantity related to 
the knotted-ness  of the vorticity lines \citep{moffatt1969degree}. 
It is also a measure of the breaking of parity invariance (mirror symmetry), with parity invariant fields being non-helical. 
Like the energy, helicity is injected in the forcing scales, cascades by the nonlinearities to the smaller scales where
it is balanced by the helicity dissipation caused by the viscous forces. This balance leads to the relation:
\beq
\left\langle{\bf F \cdot w }  \right\rangle_{_T}= \nu\left\langle  {\bf w \cdot \nabla \times w } \right\rangle_{_T}\equiv \epsilon_{_H},
\eeq
where $\epsilon_{_H}$ is the helicity dissipation rate.

The distribution of the two invariants among scales, and their transfer across scales is probably best described  through the Fourier transform of the fields that we here define as
\beq
{\bf \tilde{u}}_{\bf k}(t) =  \frac{1}{(2\pi L)^{3}}\int e^{-i\bf kx} {\bf u} d{\bf x}^3, \quad   
       {\bf u} (t,{\bf x})  = \sum_{\bf k} e^{+i\bf kx} {\bf \tilde{ u}_k} 
\eeq
and similar for the vorticity field ${\bf w}$ where ${\bf \tilde{w}_k}= i{\bf k\times \tilde{u}_k}$. 
The energy and helicity spectra are defined as
\beq
E(k) = \frac{1}{2}\sum_{k\le |{\bf q}|<k+1} |\tilde{\bf u}_{\bf q}|^2, \quad 
H(k) = \frac{1}{2}\sum_{k\le |{\bf q}|<k+1} \tilde{\bf u}_{\bf q}\cdot \tilde{\bf w}_{\bf q}^*
\eeq
where  $k$ a positive integer.
They express the distribution of the conserved quantities, $E$ and $H$, in wavenumber (and thus also scale) space.
 
The magnitude of the two cascades is measured by the energy and helicity fluxes that are denoted as 
$\Pi_{_E}(k)$ and  $\Pi_{_H}(k)$ respectably. They express the rate that the nonlinearities 
transfer energy and helicity from the set of wavenumbers $\bf q$ that satisfy ${| \bf q|} \le k$ 
to all larger wavenumbers.
Their steady state value is defined as:
\beq
\Pi_{_E}(k) \equiv - \left\langle {\bf u}^<_k{\bf \cdot ( u\times w ) } \right\rangle_{_T}, \quad 
\Pi_{_H}(k) \equiv - \left\langle {\bf w}^<_k{\bf \cdot ( u\times w ) } \right\rangle_{_T}, 
\eeq  
where $\bf u^<({\bf x})$ is the velocity field ${\bf u}$ filtered so that all the Fourier modes ${\bf q}$ with $|{\bf q}|> k$ are removed (see \cite{frisch1995turbulence}):
\beq
{\bf u}^<_k({\bf x})  = \sum_{{|\bf q|}<k} e^{+i\bf qx} {\bf \tilde{ u}_q}, \qquad
{\bf w}^<_k({\bf x})  = \sum_{{|\bf q|}<k} e^{+i\bf qx} {\bf \tilde{ w}_q}\,.
\eeq
In the limit $k\to \infty$ the fields $\bf u^<({\bf x})$ take their unfiltered value
$\lim_{k\to \infty} {\bf u^<}_k = {\bf u}$ and
two fluxes become zero $\Pi_{_E}(\infty)=\Pi_{_H}(\infty)=0$ expressing the conservation of energy and helicity by the nonlinearities.
Positive values of $\Pi_{_E}$ imply that energy cascades forward   to the large wavenumbers, while 
negative values of $\Pi_{_E}$ imply that energy cascades inversely to the small wavenumbers.
More care needs to be taken for the helicity because it is a non-sign-definite quantity.
Positive values of $\Pi_{_H}$ imply that the non-linearities decrease helicity in the large scales 
and increase helicity in the small scales. If the helicity is positive at all scales this can be interpreted as transfer of helicity from large scales to small, and thus a forward cascade. 
If however the helicity is negative at all scales the large scale helicity will increase 
in absolute value at the large scales and thus positive flux can be interpreted as transfer of negative helicity from small scales to large, and thus an inverse cascade. It is harder to give an interpretation in terms of a cascade when the helicity is not of the same sign at all 
scales, and perhaps such an interpretation in terms of a cascade should be avoided. Nonetheless,
the definition of the helicity flux is still well defined and its interpretation as the rate helicity is changing  due to the non-linearities 
inside a Fourier-space sphere  of a given radius is still valid.

Energy and helicity fluxes can be viewed as the cumulative result of a large network of 
triadic interactions of Fourier modes whose wave-vectors $\bf k$ form a triangle.
These triadic interactions allow the exchange of energy and helicity  between the three involved modes 
while conserving their sum.
They then comprise the building blocks of turbulence since their cumulative 
effect allows the transport of energy and helicity across scales leading to the turbulent cascade.
In three dimensions, the three components of the Fourier modes $\bf \tilde{u}_k $
satisfy the incompressibility condition $\bf \tilde{u}_k \cdot k=0 $ leaving two independent complex 
amplitudes. Therefore each Fourier mode can be further decomposed in two modes.  
From all possible basis that a Fourier mode of an incompressible field can be decomposed
the most fruitful perhaps has been that of the decomposition to two helical modes (see \cite{Lesieur72,Constantin1988,cambon1989spectral,waleffe1992nature}):
\beq
{\bf \tilde{u}_k} = \tilde{u}_{\bf k}^{+} {\bf h}_{\bf k}^{+} 
                  + \tilde{u}_{\bf k}^{-} {\bf h}_{\bf k}^{-}.
\eeq
The basis vectors ${\bf h}_{\bf k}^{+},{\bf h}_{\bf k}^{-}$ are 
\beq
{\bf h}^s_{\bf k}=\frac{\bf e_z \times k }{\bf \sqrt{2}|e_z \times k|} 
  +  i s \frac{\bf k\times (e_z \times k) }{\bf \sqrt{2}|k\times (e_z \times k)|}
\eeq
for ${\bf e_z \times k }\ne0$ while ${\bf h}^s_{\bf k}=({\bf e_x } + is {\bf e_y })/\sqrt{2}$ for $\bf k$ parallel to $\bf e_z $. Here $\bf e_x,e_y,e_z$ are three orthogonal unit vectors. 
The sign index $s=\pm1$ indicates the sign of the helicity of ${\bf h}^s_{\bf k}$.   
The basis vectors ${\bf h}^s_{\bf k}$ are unit norm eigenfunctions of the curl operator in Fourier space such that $i{\bf k \times h}^s_{\bf k} = s |{\bf k}| {\bf h}^s_{\bf k}$ 
and satisfy ${\bf h}^s_{\bf k} \cdot {\bf h}^s_{\bf k} =0$ and  ${\bf h}^s_{\bf k} \cdot {\bf h}^{-s}_{\bf k} =1$. They thus form a complete base for incompressible vector fields.
The velocity field for each Fourier mode $\bf k$ is then determined by the two scalar complex functions $\tilde{u}^s_{\bf k}={\bf \tilde{u}_k\cdot h}^{-s}_{\bf k}$.

\begin{figure}
  \centerline{
   \includegraphics[width=12cm]{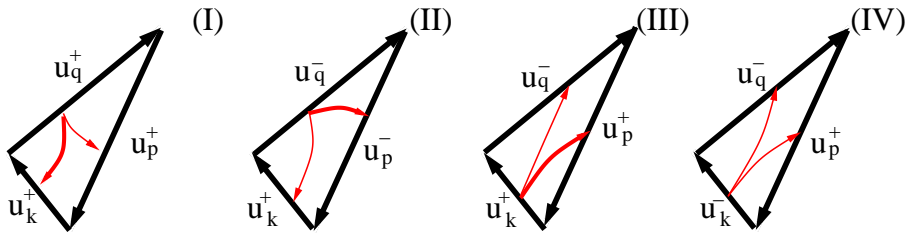}}
  \caption{ Transfer of energy in isolated in four triadic interactions between different helical modes based on \citep{waleffe1992nature},
            the remaining four interactions are obtained by interchanging the $\pm$ indexes while keeping the same direction of the flux.
            The thickness of the arrows indicates the magnitude of the transfer of energy.} 
\label{fig:triad}
\end{figure}

This decomposition was first proposed by \cite{Lesieur72}.  
Since then it has been used in several theoretical and numerical investigations in turbulence theory. 
It was also used by \cite{Constantin1988}, to study organized Beltrami hierarchies in a systematic fashion
and by \cite{cambon1989spectral} to derive an eddy damped quasi-normal Markovian model for rotating turbulence. 
In a seminal paper \citep{waleffe1992nature} considered individual triadic interactions of helical modes. 
In such isolated interactions he showed that the lowest $k$ helical mode is unstable when larger $k$ modes have helicities of opposite signs and thus, he argued, it 
can be interpreted as a mechanism to transfer energy to smaller scales. In all other cases the medium wavenumber is unstable and thus there was transfer to both large and small scales. 
For the particular case that all three modes are of the same helicity most of the transfer of energy is to the smallest wave number, and thus 
energy is transferred to the large scales. 
A schematic representation of the direction of energy transfers obtained in \citep{waleffe1992nature} is shown in figure \ref{fig:triad} 
where the magnitude of the cascade is indicated by the thickness of the arrows.
Under the assumption that the statistical behavior of the flow is controlled by the stability characteristics of these isolated triads
(referred to as the `instability assumption') \cite{waleffe1992nature} was able to draw conclusions for the direction of energy cascade in full Navier-Stokes equations \ref{NS}.
Of course a large network of triadic interactions as is the Navier-Stokes equation can behave differently from a collection of
isolated triads as has been noted recently by \cite{Moffatt2014note} and care needs to be taken when interpreting the results.

Returning to real space the velocity field can be written as 
\beq
{\bf u}(t,{\bf x})={\bf u}^+(t,{\bf x})+{\bf u}^-(t,{\bf x})
\quad
\mathrm{where}
\quad
{\bf u}^s (t,{\bf x}) =\mathbb{P}^s[{\bf u}] 
\label{helf}
\eeq
and $\mathbb{P}^s$ stand for the projection operators $\mathbb{P}^s$ of  real fields ${\bf g(x)}$ to the two different bases defined as
\beq
{\bf g}^s \equiv \mathbb{P}^s[{\bf g}] \equiv \sum_{\bf k} e^{i \bf k\cdot x} {\bf h}^s_{\bf k} ({\bf \tilde{g}_k\cdot h}^{-s}_{\bf k}) .
\eeq
Completeness and incompressibility of the bases allows us to 
write the projection operator to incompressible fields as $\mathbb{P}[{\bf g}] = \mathbb{P}^+[{\bf g}]+\mathbb{P}^-[{\bf g}]$.

The energies $E^\pm$ and helicities $H^\pm$ associated to the two fields $\bf u^\pm(t,{\bf x})$ can then be defined as
\beq
E^\pm = \frac{1}{2}\sum_{\bf k} \left| \tilde{u}^\pm_{\bf k}\right|^2, \quad H^\pm =  \pm\frac{1}{2}\sum_{\bf k} |{\bf k}| \left| \tilde{u}^\pm_{\bf k} \right|^2,
\eeq
The total energy $E$ can be written as $E=E^++E^-$ and the total helicity $H$ is written as $H = H^++H^-$.
Note that with this definition $H^-$ is a non-positive quantity.
Using the helical decomposed fields ${\bf u}^\pm,{\bf w}^\pm$ the Navier-Stokes equations can then be written as
\begin{equation}
\partial_t {\bf u}^{s_1} =  \sum_{s_2,s_3}
                            \mathbb{P}^{s_1} \left[ {\bf u}^{s_2} \times {\bf w}^{s_3} \right] 
                         +  \nu \Delta  {\bf u}^{s_1} +  \mathbb{P}^{s_1}[{\bf F}]
                         \label{NSHC}
\end{equation}
The nonlinear term of the Navier-Stokes equation is now expressed as the sum of eight terms 
$\mathbb{P}^{s_1} \left[ {\bf u}^{s_2} \times {\bf w}^{s_3} \right]$ that 
correspond to all possible permutations of the signs $s_i=\pm1$ where $i=1,2,3$.
Each of these terms has different properties concerning the the evolution of the averaged quantities $E^\pm,H^\pm$.
The evolution of the quantities $E^\pm$ and $H^\pm$ can be obtained by taking the inner product of the Navier-Stokes eq. (\ref{NSHC}) with $\bf u^\pm$ and $\bf w^\pm$ respectively, and space average. 
Leading to :
\begin{equation}
\partial_t E^{s_1}=
\sum_{s_2,s_3}   
\Big\langle {\bf u}^{s_1} \cdot \left({\bf u}^{s_2} \times {\bf w}^{s_3}\right)\Big\rangle + 
                    \nu \left\langle   {\bf w}^{s_1} \cdot  {\bf w}^{s_1}  \right\rangle +
                         \left\langle   {\bf F \cdot u}^{s_1}  \right\rangle             
                         \label{EnC} 
\end{equation}
and
\begin{equation}
\partial_t H^{s_1}=
\sum_{s_2,s_3}   
\Big\langle {\bf w}^{s_1} \cdot \left({\bf u}^{s_2} \times {\bf w}^{s_3}\right)\Big\rangle + 
                    \nu \left\langle   {\bf w}^{s_1} \cdot  \Delta {\bf u}^{s_1}  \right\rangle +
                         \left\langle   {\bf F \cdot w}^{s_1}  \right\rangle              .
                         \label{HnC}
\end{equation}
It is evident from the expressions above that the nonlinear terms $\mathbb{P}^{s_1} \left[ {\bf u}^{s_2} \times {\bf w}^{s_3} \right]$ in the sum
with $s_1=s_2=s$, conserve $E^\pm$ independently ($ie$ $\langle {\bf u}^{s} \cdot \left({\bf u}^{s} \times {\bf w}^{s_3}\right)\rangle=0$) but not $H^\pm$, 
and the nonlinear terms in with $s_1=s_3=s$, conserve $H^\pm$ independently ($ie$ $\langle {\bf w}^{s} \cdot \left({\bf u}^{s_2} \times {\bf w}^{s}\right)\rangle=0$) but not $E^\pm$.
Thus only the nonlinear terms $\mathbb{P}^{s_1} \left[ {\bf u}^{s_2} \times {\bf w}^{s_3} \right]$, in \ref{NSHC} with $s_1=s_2=s_3=s=\pm1$, 
conserve all four quantities $E^\pm,H^\pm$ independently.
The terms 
$\langle {\bf u}^{s_1} \cdot \left({\bf u}^{-s_1} \times {\bf w} \right)\rangle$ 
that do not conserve the energies $E^\pm$,
and the terms $ \langle {\bf w}^{s_1} \cdot \left({\bf u} \times {\bf w}^{-s_1}\right) \rangle$
that do not conserve the helicities $H^\pm$, 
are responsible for transferring energy and helicity from one field $\bf u^+$ 
to the other $\bf u^-$ keeping the total energy and the total helicity unaltered. 

With this in mind we can decompose the energy and the helicity flux in eight {\bf partial fluxes} as:
\begin{equation}
\Pi_{_E}^{s_1,s_2,s_3}(k) =-
\left\langle   {\bf u}^{s_1<}_k\cdot \left({\bf u}^{s_2} \times {\bf w}^{s_3}\right) \right\rangle_{_T}, \quad
\Pi_{_H}^{s_1,s_2,s_3}(k) =-
\left\langle   {\bf w}^{s_1<}_k\cdot \left({\bf u}^{s_2} \times {\bf w}^{s_3}\right) \right\rangle_{_T}
\label{flux_def}
\end{equation}
where ${\bf u}^{s<}_k$ express the two helical fields ${\bf u}^{s}_k$ given in (\ref{helf}) filtered so that only Fourier modes 
inside a sphere of radius $k$ are kept. 
The total energy and helicity flux can be recovered by summing these partial fluxes:
\begin{equation}
\Pi_{_E}(k)= \sum_{s_1,s_2,s_3} \Pi_{_E}^{s_1,s_2,s_3}(k), \quad \mathrm{and} \quad 
\Pi_{_H}(k)= \sum_{s_1,s_2,s_3} \Pi_{_H}^{s_1,s_2,s_3}(k).
\label{flux_sum}
\end{equation}
From these eight energy fluxes only the four $\Pi_{_E}^{s,s,s_3}$ come from conservative terms for $E^\pm$ and we will refer to them as {\bf conservative fluxes}.
They have the property:
\beq \lim_{k\to \infty} \Pi_{_E}^{s,s,s_3}(k) =0 \,.\eeq
The remaining four partial fluxes $\Pi_{_E}^{s,-s,s_3}$ transfer energy among the two helical fields $\bf u^\pm$ and will be referred as {\bf trans-helical energy fluxes}.
These need to be added in pairs to result in conservative fluxes $\Pi_{_E}^{s_3,th}=\Pi_{_E}^{s,-s,s_3}+\Pi_{_E}^{-s,s,s_3}$. 
We will refer to $\Pi_{_E}^{s_3,th}$ as the {\bf averaged trans-helical energy flux}. 
It is this averaged trans-helical energy flux $\Pi_{_E}^{s,th}$ that has the property 
\beq \lim_{k\to \infty} \Pi_{_E}^{s_3,th}(k)=0. \label{eq:Eth} \eeq 
For the individual terms $\Pi_{_E}^{s,-s,s_3}$ the limit
$\lim_{k\to \infty} \Pi_{_E}^{s,-s,s_3}$ is not in general zero but expresses the rate $\mathcal{T}_{_E}^{s_3}$ that $E^+$ energy is transferred to $E^-$
by interacting with the field ${\bf w}^{s_3}$:
\beq 
\mathcal{T}_{_E}^{s_3} = \lim_{k\to \infty} \Pi_{_E}^{+,-,s_3}(k) =- \lim_{k\to \infty} \Pi_{_E}^{-,+,s_3}(k) \,.
\eeq
The total rate of transfer of $E^+$ energy to $E^-$ energy is $\mathcal{T}_{_E} =\mathcal{T}_{_E}^+ + \mathcal{T}_{_E}^-$.

Similarly for the helicity, the fluxes  $\Pi_{_H}^{s,s_2,s}$ come from conservative terms and satisfy \beq \lim_{k\to \infty} \Pi_{_H}^{s,s_2,s}(k)=0. \eeq
The fluxes $\Pi_{_H}^{s,s_2,-s}$ that transfer helicity from $H^+$ to $H^-$ and visa versa will be referred as {\bf trans-helical helicity fluxes}.
They need to be paired to the {\bf averaged trans-helical helicity flux} $\Pi_{_H}^{s_2,th}=\Pi_{_E}^{s,s_2,-s}+\Pi_{_E}^{-s,s_2,s}$ to take a conservative form:
\beq 
\lim_{k\to \infty} \Pi_{_H}^{s_2,th}(k) =0 \label{eq:Hth} .
\eeq
Due to the negative sign of $H^-$ an increase of $H^-$ by the nonlinear terms in absolute value implies an equal increase in $H^+$ so that 
total helicity is conserved. The total rate of generation $\mathcal{G}_{_H}^s$ of $H^+$ helicity (equal to the rate of generation of $|H^-|$) 
through the interaction with the velocity fields ${\bf u}^s$ is defined as
\beq \mathcal{G}_{_H}^s = \lim_{k\to \infty} \Pi_{_H}^{-,s,+} = -\lim_{k\to \infty} \Pi_{_H}^{+,s,-}.\eeq
The total generation rate of $H^+$ (and thus $H^-$) is then given by $\mathcal{G}_{_H} =\mathcal{G}_{_H}^+ + \mathcal{G}_{_H}^-$.

This decomposition of the fluxes allows to study the role of different classes of interactions in a turbulent flow. 
It thus provides a way to make some contact (but not complete) with the predictions obtained from the analysis of 
individual isolated interactions in \cite{waleffe1992nature}. 
With the present description although we can link the $\Pi_{_E}^{s,s,s}$ and $\Pi_{_H}^{s,s,s}$ fluxes with the same helicity type of interactions (type $I$ depicted in figure \ref{fig:triad}).
However, we cannot we cannot link the remaining fluxes with the other types of interactions ( types II,III,IV in figure \ref{fig:triad}) 
because the calculation of the fluxes $\Pi_{_E}^{s_1,s_2,s_3}$ and $\Pi_{_H}^{s_1,s_2,s_3}$ only provides information about the direction of the cascade
and does not allow us to distinguish between the magnitude of the three wave-vectors involved in the interactions as was done in \cite{waleffe1992nature}.
Such a comparison could be made possible to some extend by considering shell-to-shell energy transfers \citep{Alexakis2005,verma2005local,Mininni2006large} that is not attempted here. 
Furthermore, 
the flux  $\Pi_{_E}^{s_1,s_2,s_3}$ expresses the flux of energy $E^{s_1}$ to $E^{s_2}$ due to the interaction 
with the field ${\bf w}^{s_3}$ that acts as a `catalysts'. Similarly $\Pi_{_H}^{s_1,s_2,s_3}$ is flux of Helicity $H^{s_1}$ to $H^{s_3}$ 
due to the interaction with the field ${\bf u}^{s_2}$ that acts as a `catalysts' for the helicity transfer and 
these fluxes represent a cumulative effect of all involved triads. 
This differs from the analysis of individual triadic interactions that is treated as a closed system and the exchange of energy and helicity
between all three modes is considered. 

An alternative approach in studying the effect of different types of interactions 
has been examined recently by \cite{Biferale2012prl,Biferale2013jfm} where
high resolution numerical simulations were carried out keeping only helicity modes of one sign. 
These interactions are the ones that suggest an inverse transfer of energy to the large scales. The authors thus solved for
\begin{equation}
\partial_t {\bf u^+}  = \mathbb{P^+}  \left[ {\bf u^+ \times w^+} \right] + \nu \Delta {\bf u^+}   + {\bf F^+}.
\label{eq:bif}
\end{equation}
This system conserves both $E^+$ and $H^+$ (while $E^-$ and $H^-$ are absent) which in this case are both positive definite quantities. 
It leads to an inverse cascade of energy and a forward cascade of helicity.
Following  the arguments of \cite{fjortoft1953changes} for two-dimensional turbulence, 
for the system given in eq.  \ref{eq:bif} we can consider the transfer of energy and helicity 
among spherical shells of radius $k_n=r^nk_0$ for some $r>1$. 
The conservation of the total energy $E^+$ and total helicity $H^+=kE$ for the transfer from the set of wavenumbers with 
${|\bf k|}=k_n$ to a set of wavenumbers with ${|\bf k|}=rk_n$ and ${|\bf k|}=k_n/r$. 
then imposes that the fraction $r/(r+1)>1/2$ of energy $E_n^+$ is transferred to the large scales $k_n/r$ while a smaller part of the energy 
$1/(r+1)<1/2$ is transferred to the small scales $k_nr$ and contrary for the Helicity $H^+$.
This argument would no longer be valid if $H^+$ was not a sign definite quantity. Note that \cite{fjortoft1953changes} arguments only consider 
the presence of conserved quantities and are independent of the exact form of the nonlinear interactions. 
A constant flux of energy to the large scales leads to a Kolmogorov energy spectrum $E_k^+\propto k^{-5/3}$
while following the same arguments for forward energy cascade of the helicity one obtains the energy
spectrum  $E_k\propto k^{-7/3}$. Both of these spectra were realized in the simulations of \cite{Biferale2012prl,Biferale2013jfm}.
More recently these investigations of decimated models of the Navier-Stokes equations were carried out further
by either explicitly eliminating only a fraction of the $\tilde{u}^-_{\bf k}$ modes \cite{sahoo2015disentangling,sahoo2015role}
or by suppressing the negative helicity modes by dynamical forcing function \cite{stepanov2015hindered}. 
The work of \cite{sahoo2015disentangling,sahoo2015role} showed that the inverse cascade of energy appears only when
all the all the negative helical modes are removed.

In this work a different direction is followed. Instead of removing part of the interactions from the Navier-Stokes,
all classes of interactions are kept but their individual effect is followed by monitoring the decomposed fluxes. 
To this end large scale numerical simulations are performed both in the absence of global helicity and in its presence.  
Details of the simulations are given in the next section, while the result from the fluxes decomposition are given in section
\ref{sec:fluxes}. Conclusions are drawn in the last section.

\section{Simulations}\label{sec:num}                      
 
 To unfold the implications of the proposed decompositions in the previous section we performed 
numerical simulations of the Navier-Stokes equations in a triple periodic cubic domain of size $2\pi$
at resolution $1536^3$. The simulations were performed using the pseudospectral {\sc Ghost}-code \citep{mininni2011hybrid}
with a fourth order Runge-Kutta method for the time advancement and 2/3 rule for de-aliasing.
The flow was forced by a mechanical forcing $\bf F$ that consisted only of 
Fourier modes with wavenumbers $\bf q$ such that $k_f\le |{\bf \bf q}|\le k_f+1$ with $k_f=4$. 
This relative high wave number of the forcing is chosen so that not only the forward cascade is studied but
the behavior of the flow at scales larger than the forcing are also examined.
The amplitude of the forcing was fixed at unity $\| {\bf F}\|=1$ and the phases of the Fourier modes were changed
randomly at fixed time intervals $\tau_f=0.1$. Two different forcing functions were considered; 
in the first the Fourier modes of the forcing were not helical (so $\|\mathbb{P}^+[{\bf F}]\|=\|\mathbb{P}^-[{\bf F}]\|$), 
while in the second each Fourier mode was fully-helical with positive helicity (ie ${\bf F}=\mathbb{P}^+[{\bf F}]$ and $\mathbb{P}^-[{\bf F}]=0$ at each instant of time).
All the parameters of the runs and the basic observables are given in
table \ref{tab:param}. To gain computational time, the runs were started using as initial conditions the results from runs with smaller $Re$ 
(and smaller grid)  and were continued for twelve turnover times ($\tau_u=1/\|{\bf u}\|k_f$) after the first peak of energy dissipation appeared. 
 
 \begin{table}
  \begin{center}
  \begin{tabular}{cccccccccc}
      case   &$ \quad  \nu, \quad $  &$\quad  k_f, \quad$ & $ \quad\|\bf F\|, \quad$ & $\quad \tau_f,\quad $ & $ \quad \| {\bf u} \|, \quad$ & $ \quad\epsilon,\quad$ &  $Re\equiv \frac{\| {\bf u} \|}{k_f \nu}, $ & $k_{m}\ell_\nu,$ &   $H/Ek_f$, \\[3pt]
      \hline
 Non-helical & 0.0002 & 4     &    1.0    & $0.1$  &  0.948         &    0.194    &  1185    &  1.292        &   0.009     \\
 Helical     & 0.0002 & 4     &    1.0    & $0.1$  &  1.072         &    0.191    &  1340    &  1.373        &   0.859     \\
  \end{tabular}
  \caption{Parameters of the numerical simulations for the helical and the non-helical case. In both simulations $N=1536^3.$
  $k_m=N/3$ is the maximum wavenumber, and $\ell_\nu=(\nu^3/\epsilon)^{1/4}$ is the Kolmogorov length-scale. }
  \label{tab:param}
  \end{center}
\end{table}

\begin{figure}
  \centerline{
  \includegraphics[width=7cm]{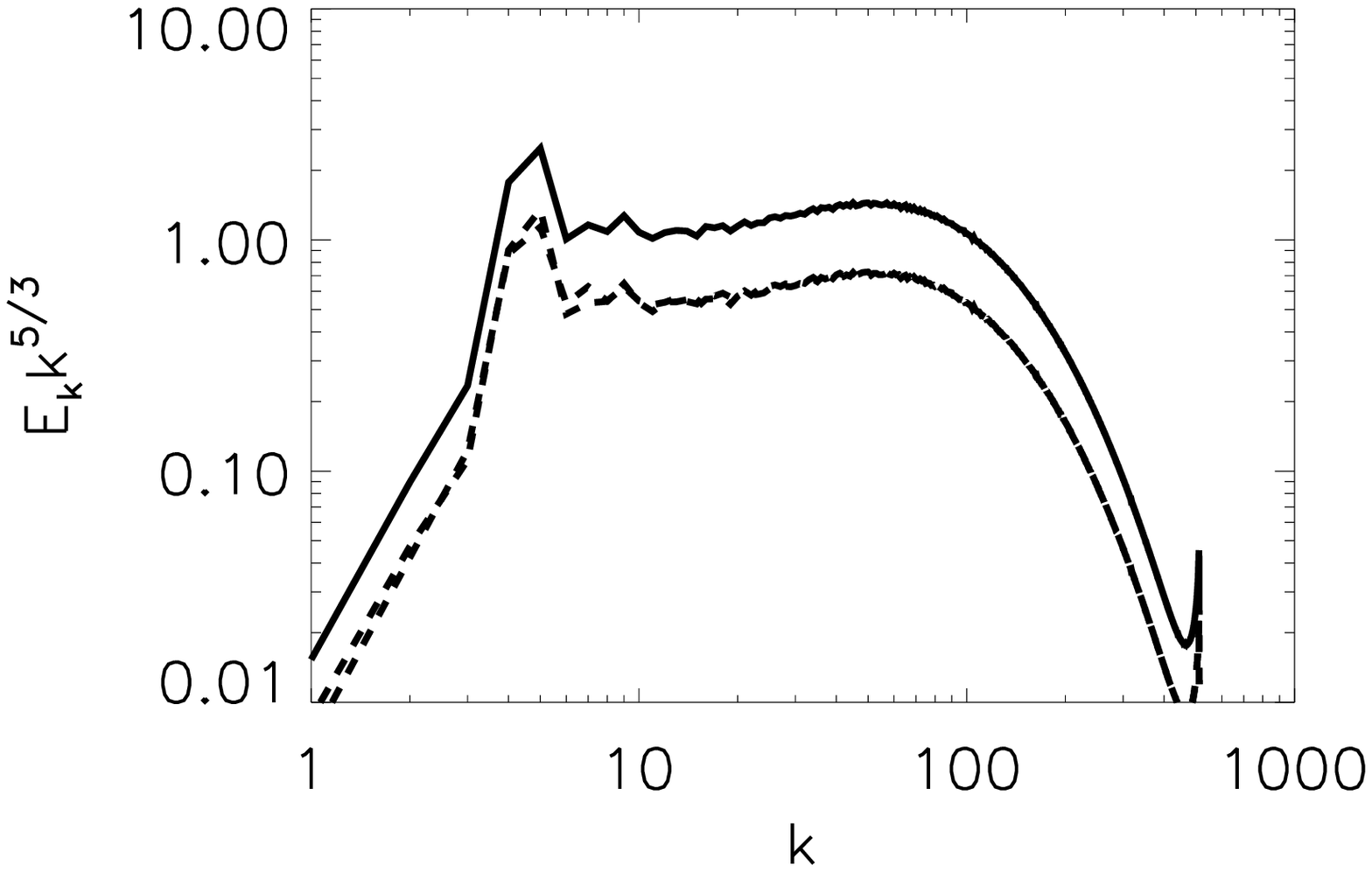}
  \includegraphics[width=7cm]{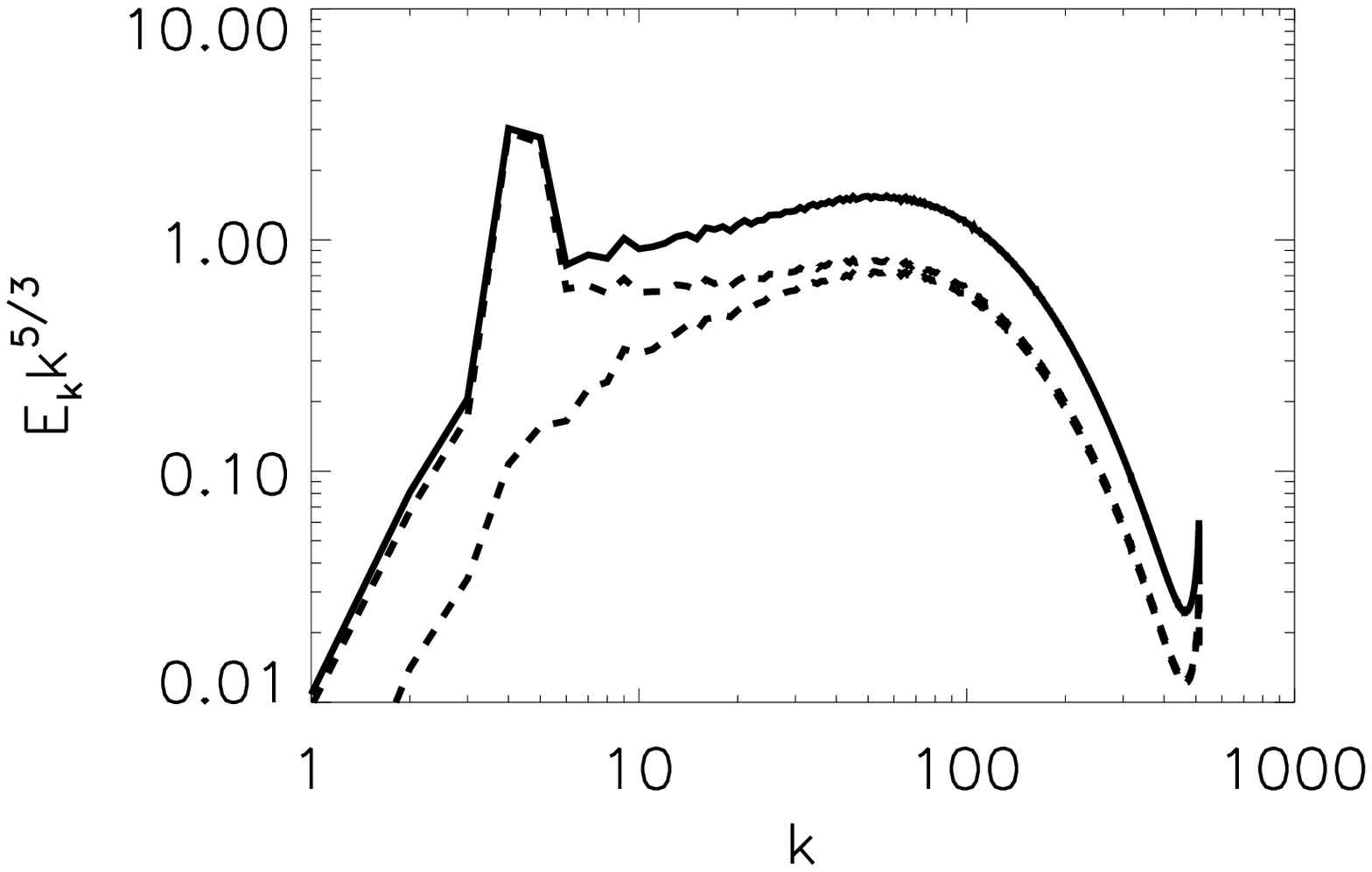}}
  \caption{ Energy spectra compensated by $k^5/3$
  for the non-helical case (left) and the helical case (right). The solid line shows the total energy spectrum $E_k=E^+_k+E^-_k$, and the dashed lines
  show the two spectra $E^\pm_k$. For the non-helical case the spectra $E^\pm_k$ are indistinguishable while for the helical case $E^+_k$ is significantly larger as small $k$ 
  but reaches equipartition with $E^-_k$ at large $k$. Both spectra show a bottleneck at small $k$ with it being more pronounced for the helical case.}
\label{fig:spec}
\end{figure}

 The resulting energy spectra of these runs compensated by $k^{-5/3}$ are shown in figure \ref{fig:spec} with a solid black line.
 The spectra show a close to  $k^{-5/3}$ behavior although a large bottleneck makes the spectra to deviate 
 from this value. The bottleneck effect although not fully understood it is very well documented 
 (\cite{herring1982comparative,falkovich1994bottleneck,lohse1995bottleneck,martinez1997energy,kurien2004cascade}) 
 and it is argued to be related to the quenching of local interactions close to the dissipative scales that leads to `pile-up' 
 of energy at these scales. It is stronger in the helical case that dominates most of the spectrum.
 
 The dashed lines in figure \ref{fig:spec} show the spectra $E^\pm_k$ defined as
\beq
E^\pm_k = \frac{1}{2}\sum_{k\le |{\bf q}|<k+1} |\tilde{\bf u^\pm}_{\bf q}|^2. \quad 
\eeq
For the non-helical case (on the left panel) the  two spectra are indistinguishable and
the two fields $\bf u^+$ and $\bf u^-$ have identical statistics. For the helical case (on the right panel) the spectrum $E_k^+$ (top dashed line) for the positive helical field 
dominates at the large scales. This is expected since the forcing injects energy only at the $\bf u^+$ modes. 
It is also worth noting that the $E_k^+$ spectrum shows a more clear $k^{-5/3}$ scaling. The  spectrum $E_k^-$ for the negative helical 
field is sub-dominant at large scales scales but increases and reaches equipartition with $E_k^+$ at large wave-numbers,
restoring parity invariance at small scales. 
 
\section{Fluxes}\label{sec:fluxes}               

The results of these simulations were used to calculate the partial fluxes defined in eq. \ref{flux_def}.
The calculation was performed at run-time at frequent time intervals and the results were time averaged at the end.
The calculation was performed in the following way. 
At each output time the fields ${\bf u }^{s_1}$, ${\bf w}^{s_1}$ and 
the four nonlinear terms ${\bf u }^{s_2} \times {\bf w}^{s_3}$ were calculated.
Then the inner product in eq. \ref{flux_def} with ${\bf u }^{s_1<}$ and ${\bf w}^{s_1<}$ was obtained by filtering
${\bf u }^{s_1}$ and ${\bf w}^{s_1}$. This procedure is eight times as costly as one Runge-Kutta time step  
but since the fluxes were not calculated every time step it did not lead to a significant slow down of the code.
The results were finally averaged over the steady state and are presented in the subsections that follow.

\subsection{Energy Fluxes}                    

\begin{figure}
\centerline{
\includegraphics[width=7cm]{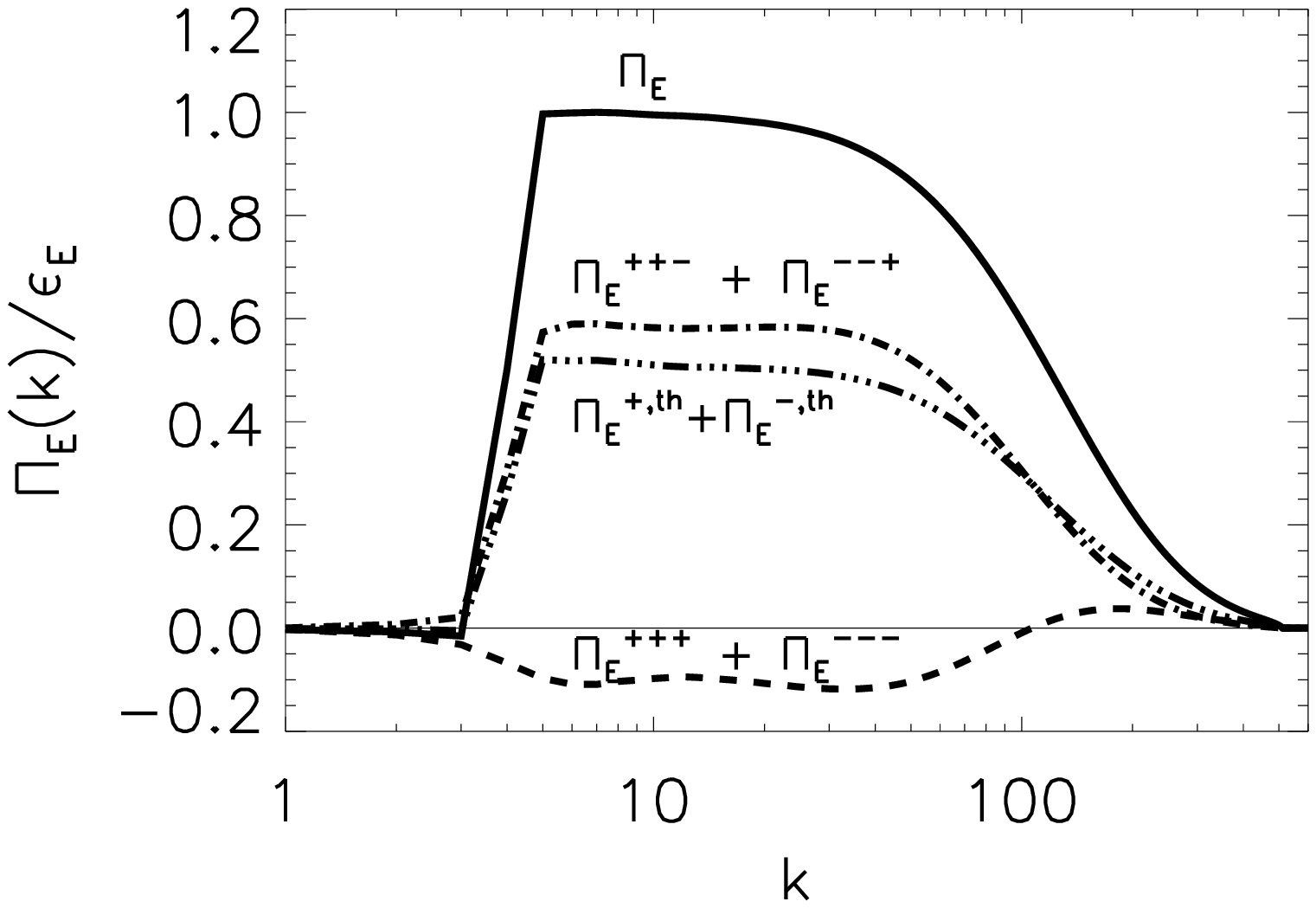}
\includegraphics[width=7cm]{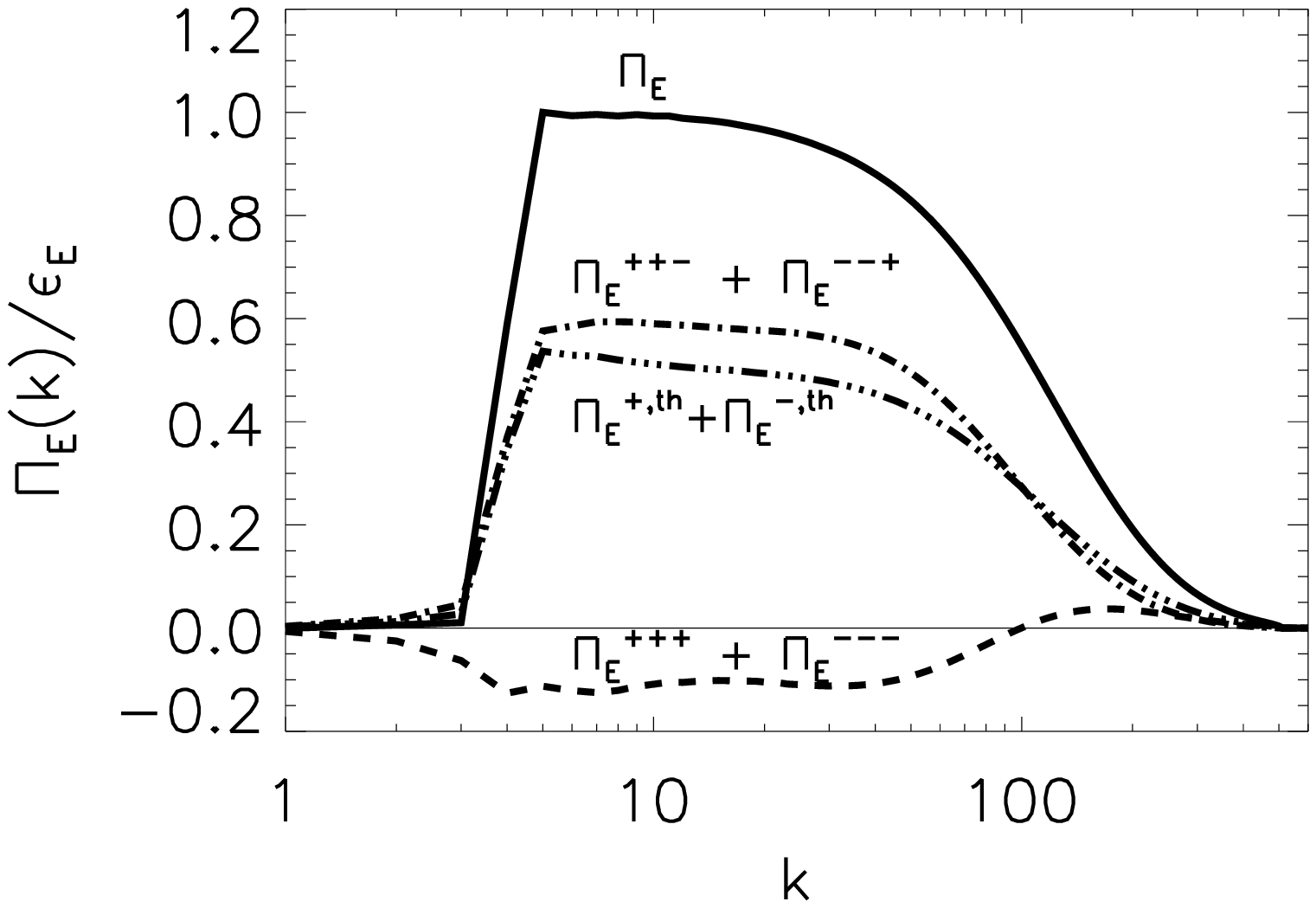}}
\caption{ Total energy flux $\Pi_E$ (solid line) and the symmetrized energy fluxes 
$\Pi_E^{+++}+\Pi_E^{---}$ (dashed line), 
$\Pi_E^{++-}+\Pi_E^{--+}$ (dash dot line),
$\Pi_E^{+,th}+\Pi_E^{-,th}$ (dash dot dot dot line),
for the non-helical (left) and the helical (right) case. }
\label{fig:flux_1}
\end{figure}

First the energy fluxes are examined.
Figure \ref{fig:flux_1} shows with a solid line the total energy flux 
for the non-helical run on the left panel and for the helical run on the right panel.
The partial fluxes $\Pi_{_E}^{+++}+\Pi_{_E}^{---}$ are shown with a dashed line,
$\Pi_{_E}^{++-}+\Pi_{_E}^{--+}$ are shown with a dash-dot line, while the averaged trans-helical fluxes
$\Pi_{_E}^{+,th}+\Pi_{_E}^{-,th}$ are shown with a dash-dot-dot-dot line.
The fluxes have been summed (symmetrized) over the two signs for clarity. 
This has no effect on the non-helical flow for which the two fields have 
identical statistical properties, but it does have an effect for the helical 
run that we analyze further in what follows. 

Three striking points can be observed from figure \ref{fig:flux_1}.
First, the three symmetrized partial fluxes shown in this figure
are approximately constant in the inertial range. This is not a trivial result
as conservation of energy implies constancy of only the total energy flux.
A second observation is that the two simulations despite having a different distribution
of energy among helical modes, have identical symmetrized partial fluxes.  
Finally, and perhaps most striking is the fact that the fluxes $\Pi_{_E}^{+++}+\Pi_{_E}^{---}$ are constant and negative. 
This implies  that in turbulence, hidden inside the forward cascade of the total energy there is a process that transfers energy back to the large scales in a constant 
rate across scales. 
This is in agreement with the prediction of \cite{waleffe1992nature} that same helicity interactions transfer energy to large scales.
The amplitude of this inverse flux  is approximately 10\% of the total flux. 
This percentage is the same for both simulations and is possibly universal.
The $\Pi_E^{++-}+\Pi_E^{--+}$ and $\Pi_E^{+,th}+\Pi_E^{-,th}$ fluxes are almost equal and positive at all wave numbers 
in the inertial range with a slight excess of flux for the $\Pi_{_E}^{++-}+\Pi_{_E}^{--+}$ over  $\Pi_{_E}^{+,th}+\Pi_{_E}^{-,th}$. 
These fluxes are responsible for the total forward flux of energy.
At scales larger than the forcing scale $\Pi_{_E}^{+++}+\Pi_{_E}^{---}$ remains negative and it is balanced
by the other two fluxes $\Pi_E^{++-}+\Pi_E^{--+}$ and $\Pi_E^{+,th}+\Pi_E^{-,th}$ leading to a zero total flux for $k<k_f$.
Large scales thus reach an equilibrium by receiving energy from the small scales by 
$\mathbb{P}^s  \left[ {\bf u}^s \times {\bf w}^s \right]$ interactions and losing energy to the small scales by
the remaining interactions.
At the viscous scales the fluxes $\Pi_{_E}^{+++}+\Pi_{_E}^{---}$ change sign and become positive. 
This is because at these scales the energy spectrum is even steeper than the $k^{-7/3}$
that these interactions want to equilibrate to and thus transfer energy forward.

\begin{figure}
  \centerline{
  \includegraphics[width=7cm]{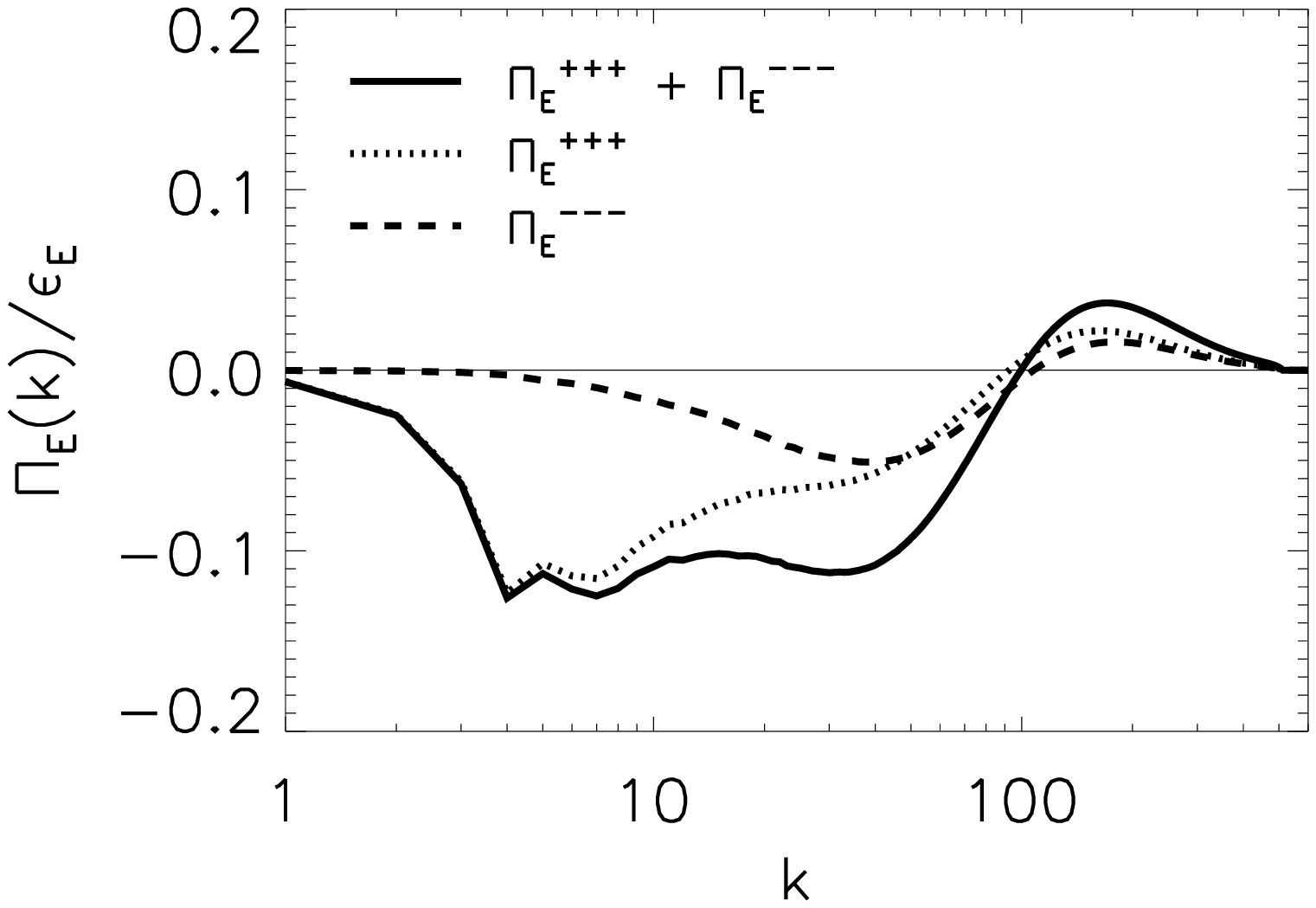}
  \includegraphics[width=7cm]{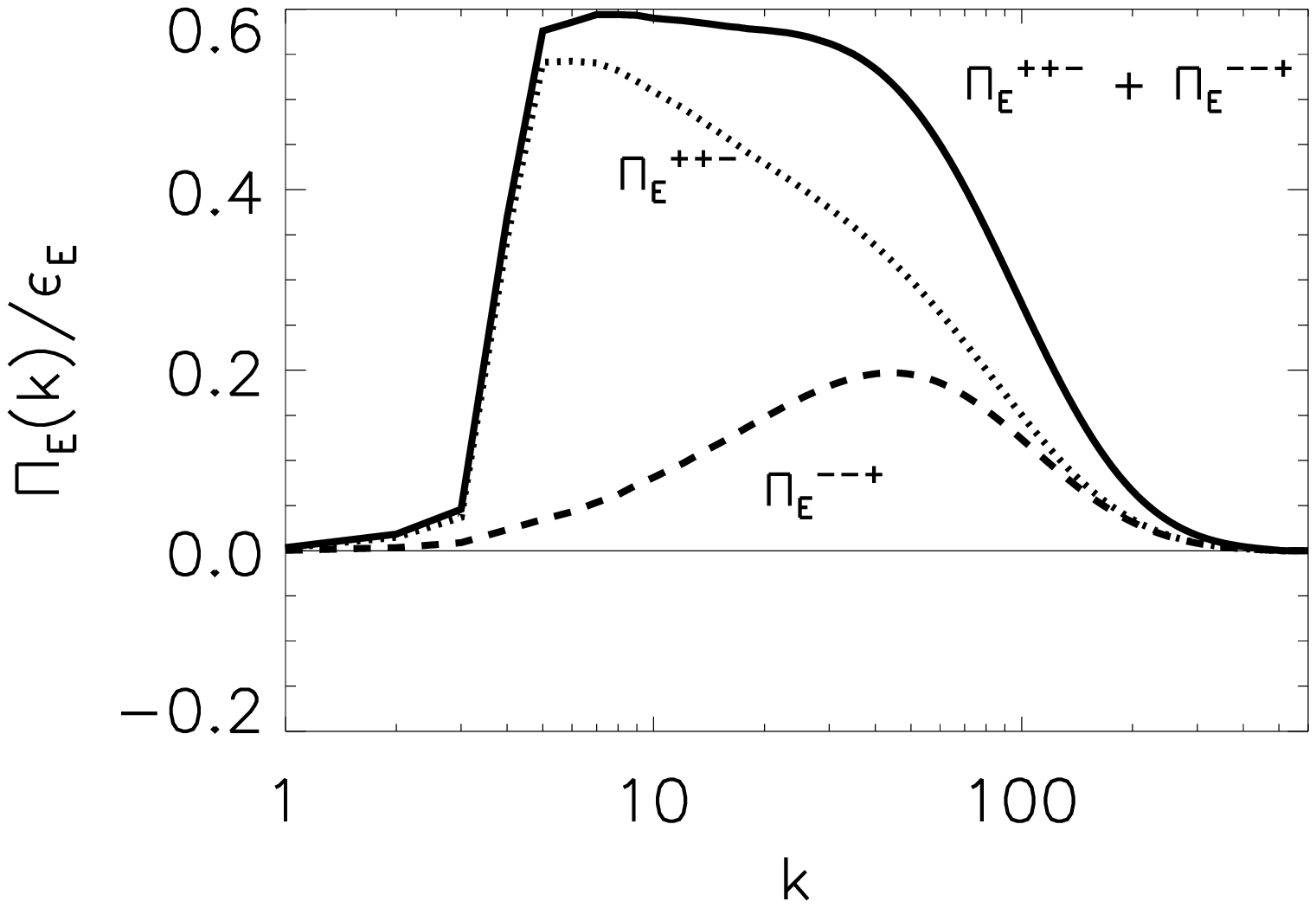}}
  \caption{ Left panel: Symmetrized energy flux $\Pi_E^{+++}+\Pi_E^{---}$ (solid line),
                        $\Pi_E^{+++}$ (doted line) and $\Pi_E^{---}$ (dashed line) for the helical run.
            Right panel:Symmetrized energy flux $\Pi_E^{++-}+\Pi_E^{--+}$ (solid line),
                        $\Pi_E^{++-}$ (doted line) and $\Pi_E^{--+}$ (dashed line) for the same run. }
\label{fig:flux_2}
\end{figure}

As discussed before, in the non-helical run the fields $u^+$ and $u^-$ have the same statistical properties
and the fluxes obey $\Pi_E^{s_1,s_2,s_3}=\Pi_E^{-s_1,-s_2,-s3}$. This is not true for the helical case
for which the $\tilde{u}^+$ modes have different distribution from the $\tilde{u}^-$ modes. To show this difference
the left panel in figure \ref{fig:flux_2} shows the symmetrized flux $\Pi_E^{+++}+\Pi_E^{---}$ along with 
the individual partial fluxes $\Pi_E^{+++}$ and $\Pi_E^{---}$. At large scales where the positively helical modes dominate
most of the inverse energy flux is driven by  $\Pi_E^{+++}$ but at smaller scales the two fluxes become equal.
A similar behavior is shown in the right panel of figure \ref{fig:flux_2} for the fluxes $\Pi_E^{++-},\Pi_E^{--+}$. 
At large scales the $\Pi_E^{++-}$ dominates because it involves two $\tilde{u}^+$ modes and 
one $\tilde{u}^-$ mode so it is stronger than $\Pi_E^{--+}$ that involves only
one $\tilde{u}^+$ mode. At small scales however that parity invariance is restored the two fluxes become equal. 

\begin{figure}
  \centerline{
  \includegraphics[width=7cm]{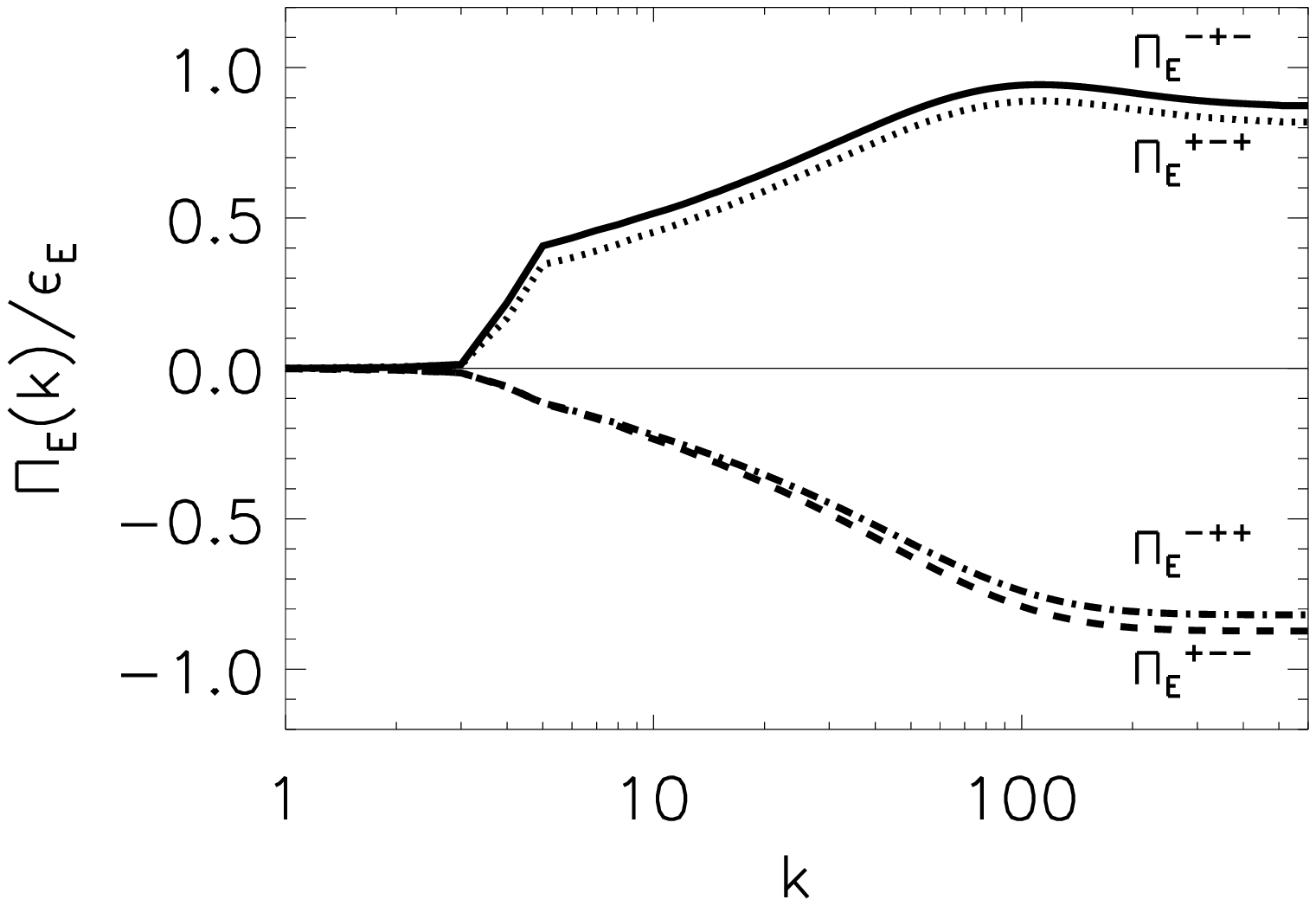}
  \includegraphics[width=7cm]{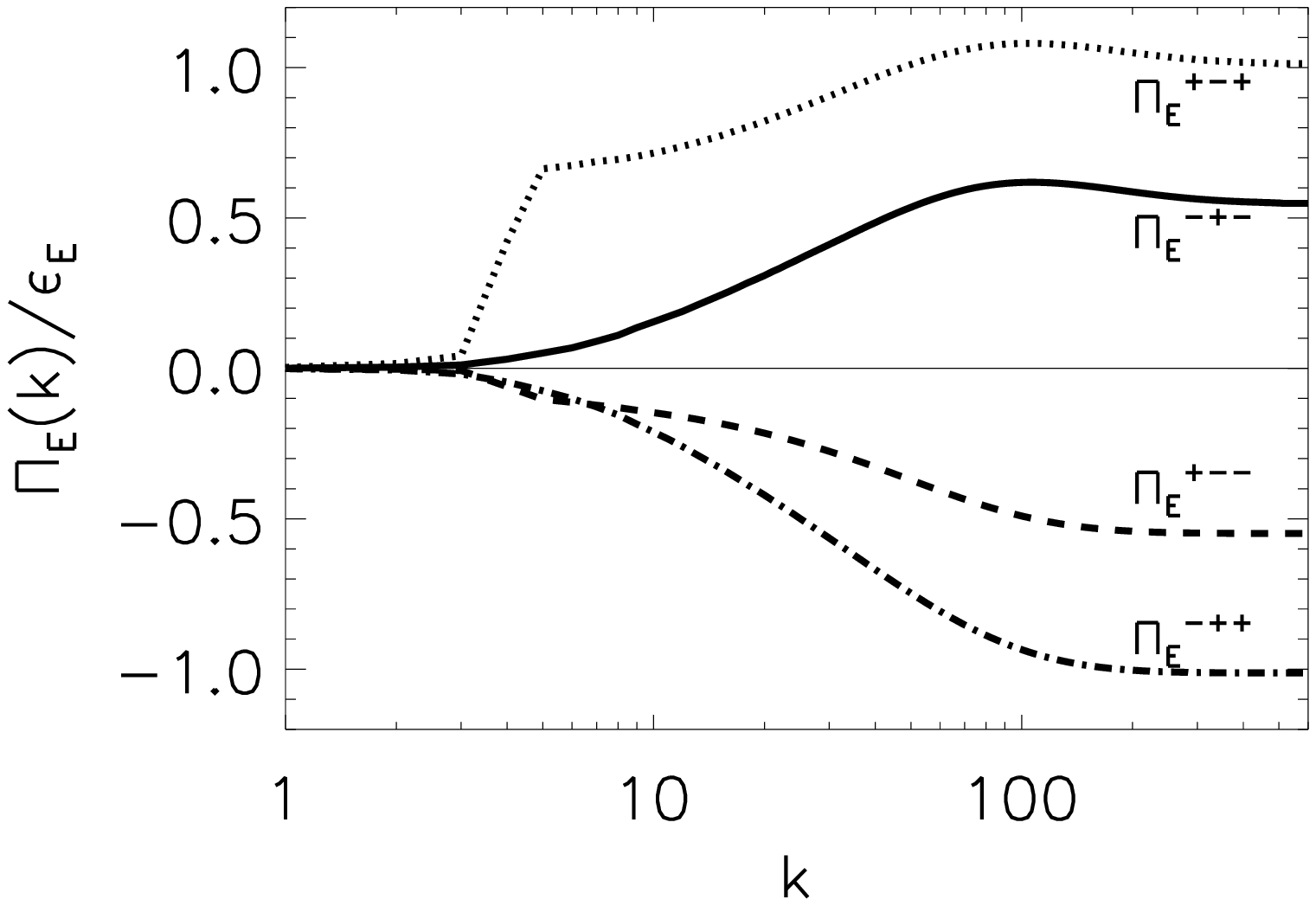}}
  \caption{ The four trans-helical energy fluxes 
  for the non-helical (left) and the helical (right) case:
  $\Pi_E^{-+-}$ (solid  line), 
  $\Pi_E^{+-+}$ (doted  line),
  $\Pi_E^{+--}$ (dashed line),
  $\Pi_E^{-++}$ (dash dot line).
  }
\label{fig:flux_3}
\end{figure}

The  trans-helical energy fluxes $\Pi_E^{-+-}$, $\Pi_E^{+-+}$, $\Pi_E^{+--}$, $\Pi_E^{-++}$  are 
plotted in figure \ref{fig:flux_3} for the non-helical case in the left panel and the helical case in the right panel.
As discussed in the introduction these fluxes originate from terms that do not conserve the individual energies $E^\pm$
but transfer energy from modes of one helicity sign to modes of the opposite helicity. 
More precisely 
$\Pi_E^{+--}$ and $\Pi_E^{+-+}$ represent the rate $E^+$ energy is transferred from the large scales to 
$E^-$ energy (at all scales) through the interaction with the $\bf w^-$ and $\bf w^+$ fields respectably. 
Similarly  $\Pi_E^{-+-}$ and $\Pi_E^{-++}$ represent the transfer rate of $E^-$ energy from the large scales to $E^-$ energy 
through the interaction with the $\bf w^-$ and $\bf w^+$ fields respectably. 
Energy conservation then implies that at $k\to \infty$ we have 
\beq
\lim_{k\to\infty}\Pi_E^{-+-}=-\lim_{k\to\infty}\Pi_E^{+--}\quad \mathrm{and} \quad \lim_{k\to\infty}\Pi_E^{-++}=-\lim_{k\to\infty}\Pi_E^{+-+}
\label{encon}
\eeq
as a direct consequence of eq. \ref{eq:Eth}.

We begin with the non-helical case. As can be seen $\Pi_E^{+-+}$ is positive at all scales. This implies
that the interactions $\mathbb{P^+}[{\bf u^- \times w^+ }]$ remove energy from the positively helical large scale modes.
On the contrary $\Pi_E^{+--}$ is negative at all scales and this implies
that the interactions $\mathbb{P^+}[{\bf u^- \times w^- }]$ increase the energy of the positively helical large scale modes.
The same conclusion can be drawn for $\Pi_E^{-+-}$ and $\Pi_E^{-++}$ for the energy of the negatively helical modes.
The end values at $k\to\infty$ of the flux $\Pi_E^{+-s}$ indicate the total rate $\mathcal{T}^{s}_{_E}$ that the
energy is transferred from the $\bf u^+$ field to the  $\bf u^-$ through the interactions $\mathbb{P^+}[{\bf u^- \times w}^s ]$.
What is observed is that
$\mathcal{T}^{+}_{_E} = \lim_{k\to\infty} \Pi_E^{+,-,+}(k) > 0$ thus this transfer removes energy from the positively helical field while
$\mathcal{T}^{-}_{_E} = \lim_{k\to\infty} \Pi_E^{+,-,-}(k) < 0$ and thus this transfer feeds with energy the modes with positive helicity.
In other words interactions with modes $\tilde{\bf w}_{\bf k}^s$ tend to transfer energy from $E^s$ to $E^{-s}$.
At the non-helical steady state the interactions with both fields $\tilde{\bf w}_{\bf k}^\pm$ reach an equilibrium with zero net transfer of energy across the two fields. 
This is realized by observing that in the  limit $k\to \infty $  we have
$\Pi_E^{-+-} \simeq \Pi_E^{+-+}$ and $\Pi_E^{-++} \simeq \Pi_E^{+--}$.

This is no longer true for the non-helical case. Although these fluxes have the same sign as in the non helical case their amplitudes are not equal. 
The interactions that involve more ($ie$ two) positive helical modes 
dominate in absolute magnitude over the interactions with more ($ie$ two) negative helical modes.
Thus the fluxes $\Pi_E^{+-+},\Pi_E^{-++}$  that lead to the transfer of energy $\mathcal{T}^{+}_{_E}$ from the positive helical modes to the negative helical modes
dominate over the fluxes  $\Pi_E^{+--},\Pi_E^{-+-}$ that display a transfer from $E^-$ to $E^+$.  
This is how parity invariance is recovered at small scales: the initial excess of $E^+$ energy leads to faster transfer $\mathcal{T}^{+}_{_E}$ from $E^+$ to $E^-$
compared to $\mathcal{T}^{-}_{_E}$ that displays a transfer in the opposite direction. 
The end values in the  limit $k\to \infty $ indicate that $\Pi_E^{-+-} \simeq \frac{1}{2} \Pi_E^{+-+}\simeq  \frac{1}{2} \epsilon_{_E}$ 
                                              and similar $\Pi_E^{-++} \simeq \frac{1}{2} \Pi_E^{+--}\simeq -\frac{1}{2} \epsilon_{_E}$.
Thus the total rate of transfer of energy from ${\bf u_k^+}$ modes to ${\bf u_k^-}$ is approximately $\mathcal{T}=\lim_{k\to\infty}(\Pi_E^{+-+}+\Pi_E^{+--}) \simeq \frac{1}{2} \epsilon_{_E}$. 
This only reflects the fact that since parity invariance is restored at small scales the two fields dissipate energy
at the same rate. In order to achieve this half of the injected energy at the ${\bf u_k^+}$ modes has to be transferred to the unforced ${\bf u_k^-}$ modes.

%
\subsection{Helicity Fluxes}                     

In this section we focus on the flux of helicity.  
The important difference between the energy and the helicity fluxes is the negative sign of $H^-$. 
For the energies $E^\pm$, the nonlinear interactions can increase $E^-$ only at the cost of decreasing $E^+$, keeping their sum the same. 
For the helicities $H^\pm$,  the negative $H^-$ can be increased in absolute value by simultaneously increasing $H^+$, 
thus generating both $H^+$ and $H^-$. This generation of $H^\pm$ is important for the sustainment of the forward energy cascade.  
As the energies $E^\pm$ are transferred to small scales the helicities $H^\pm$ that scale like $H^\pm_k = \pm k E^\pm_k$ have to increase.
This can be achieved through the interactions $\langle {\bf w}^{s} \cdot ({\bf u}^{s_2} \times {\bf w}^{-s} ) \rangle $ that do not conserve $H^\pm$ individually. 
This simultaneous generation of $H^+$ and $H^-$ is measured by the trans-helical fluxes $\Pi_{_H}^{s,s_2,-s}$. 
The remaining fluxes conserve $H^\pm$ individually and can more easily be interpreted as cascades in the following way.
The fluxes $\Pi_H^{+++}$, $\Pi_H^{+-+}$, originate from terms that conserve $H^+$ that is a positive quantity.
Thus these fluxes give a measure of the forward cascade of $H^+$ when they are positive and
of the inverse cascade of $H^+$ when they are negative. Similarly the fluxes $\Pi_H^{---}$, $\Pi_H^{-+-}$, 
originate from terms that conserve $H^-$ that is a negative quantity. Thus these fluxes can also be interpreted as measures 
of a cascade but due to the negative sign of $H^-\le0$ positive values imply an inverse cascade of $H^-$ while 
negative  values imply a forward cascade of $H^-$.

\begin{figure}
  \centerline{
  \includegraphics[width=7cm]{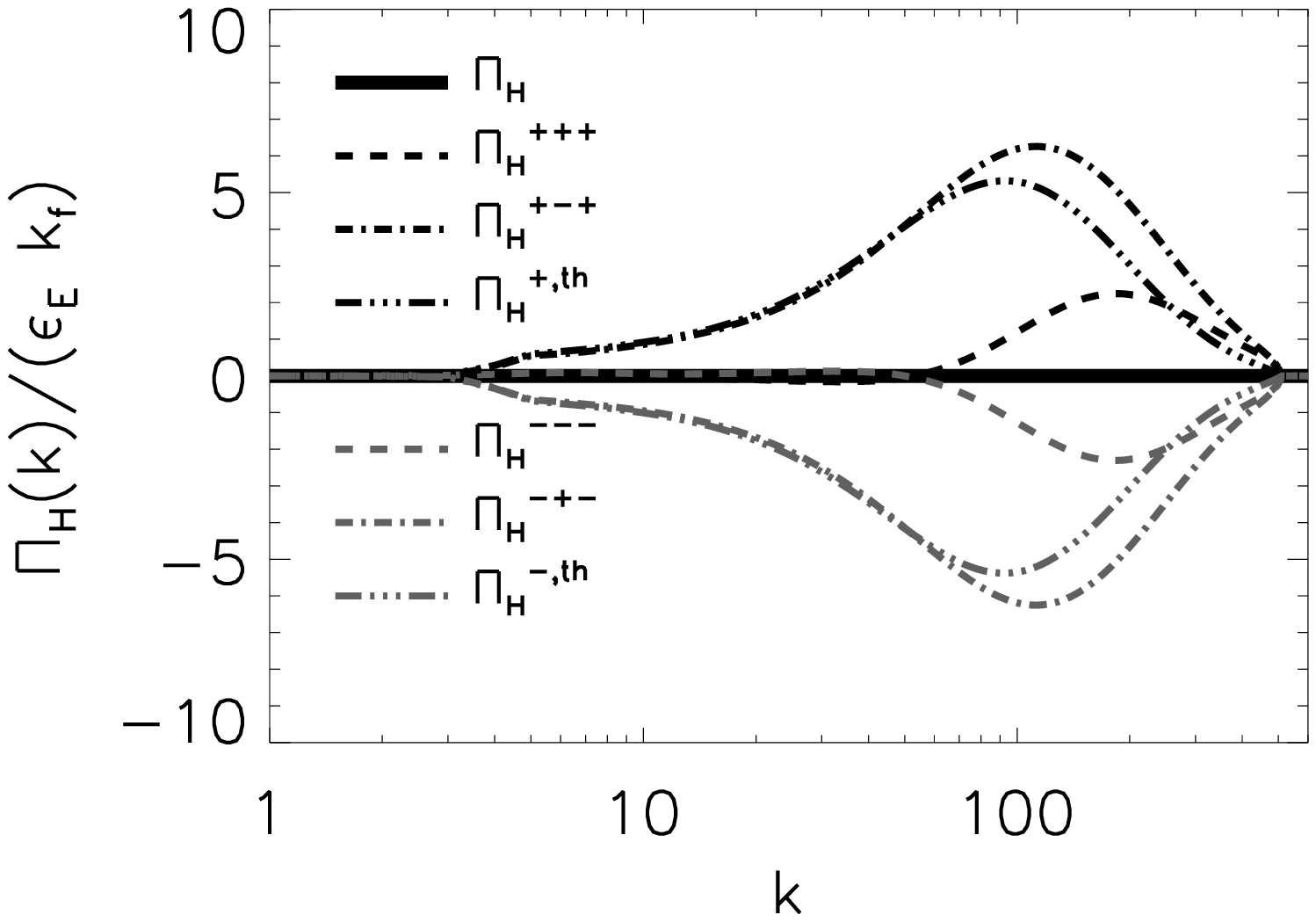}
  \includegraphics[width=7cm]{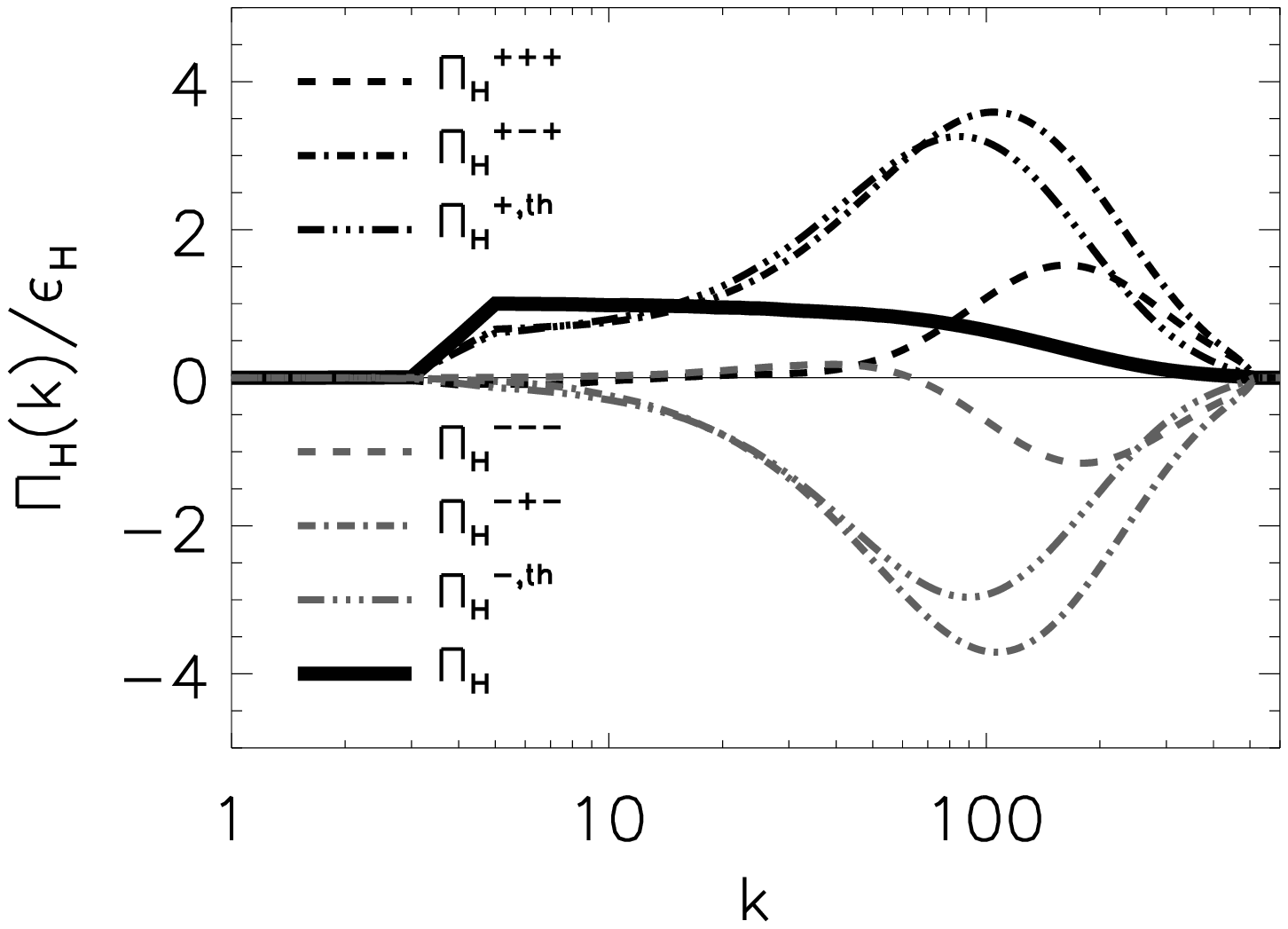}}
  \caption{ Total helicity flux $\Pi_{_H}$ (solid line) and the partial helicity fluxes 
  $\Pi_H^{+++}$, $\Pi_H^{---}$, 
  $\Pi_H^{+-+}$, $\Pi_H^{-+-}$,
  $\Pi_H^{+,th}$, $\Pi_H^{-,th}$,
  for the non-helical (left) and the helical (right) case. For the non-helical case the total helicity flux is 
  zero. }
\label{fig:flux_4}
\end{figure}

 
Figure \ref{fig:flux_4} shows with a solid line the total helicity flux 
for the non-helical run on the left panel and for the helical run on the right panel.
In the non-helical case the total helicity flux is of course zero while 
in the helical flow a constant positive flux of helicity is observed in the inertial range. 
Since the helicity of the flow is strictly positive at all scales this positive flux can be 
interpreted as a forward cascade of helicity.
%
The partial fluxes of helicity defined in eq. \ref{flux_def} are shown by the non-solid lines. 

None of the partial fluxes appear to be constant in the inertial range instead they appear to increase as the viscous scales are approached,
and then decrease again after the viscous  cut-off.
The fluxes of helicity due to same helicity interactions, $\Pi_H^{+++},\Pi_H^{---}$, are almost zero at the inertial range 
implying that they drive a weak or no cascade of helicity.  This is true both for the helical and the non-helical flow.
In the dissipation range where the spectra are much steeper $\Pi_H^{+++}$ becomes positive and $\Pi_H^{---}$ negative 
transferring thus $H^+$ and $H^-$ to the small scales.
The fluxes $\Pi_H^{+-+}$, $\Pi_H^{-+-}$, that also conserve individually $H^\pm$ are non zero but not constant. 
$\Pi_H^{+-+}$ that measures the transport of $H^+\ge0$ is positive and $\Pi_H^{-+-}$ that measures the transport of $H^-\le0$ is negative
thus the quantities $H^\pm$ are transported to the small scales {\it ie} forward cascading.  
Finally, the averaged trans-helical fluxes $\Pi_H^{+,th}$ and $\Pi_H^{-,th}$ are shown to be of the 
same amplitude as of $\Pi_H^{+-+}$, $\Pi_H^{-+-}$. The positivity of $\Pi_H^{+,th}$ implies
that the advection of the vorticity field by the $\bf u^+$ flow tends to decrease (in sign) the helicity in the large scales
while its advection by the $\bf u^-$ flow tends to increase (in sign) the helicity in the large scales. 
This phenomenon is analogous to the passive advection of magnetic field lines by helical flow 
where it is known that their stretch by a positive helical flow leads 
at a positive `twist' helicity at small scales and to a large scale negative `writhe' helicity at large scales \citep{gilbert2002magnetic,Brandenburg2005}.
This process is referred to as the stretch twist dynamo \citep{Vainshtein1972,gilbert1995stretch}.
In the helical case  due to the excess of positive helicity the fluxes are  dominated by $\Pi_H^{+-+}$ and $\Pi_H^{+,th}$ that involve 
interactions with two positive helical modes. This leads to the forward flux obtained for the total helicity.
It is worth pointing out that the values of the partial fluxes close to the dissipation scales are much larger 
than the injection values. This is due to the generation of $H^+$ and $H^-$ at the small scales by the trans-helical terms 
in such a way that their sum remains constant. This is examined in the next figure \ref{fig:flux_5} where the trans-helical fluxes 
$\Pi_H^{s,+,-s}$ and $\Pi_H^{s,-,-s}$ are shown.

\begin{figure}
  \centerline{
  \includegraphics[width=7cm]{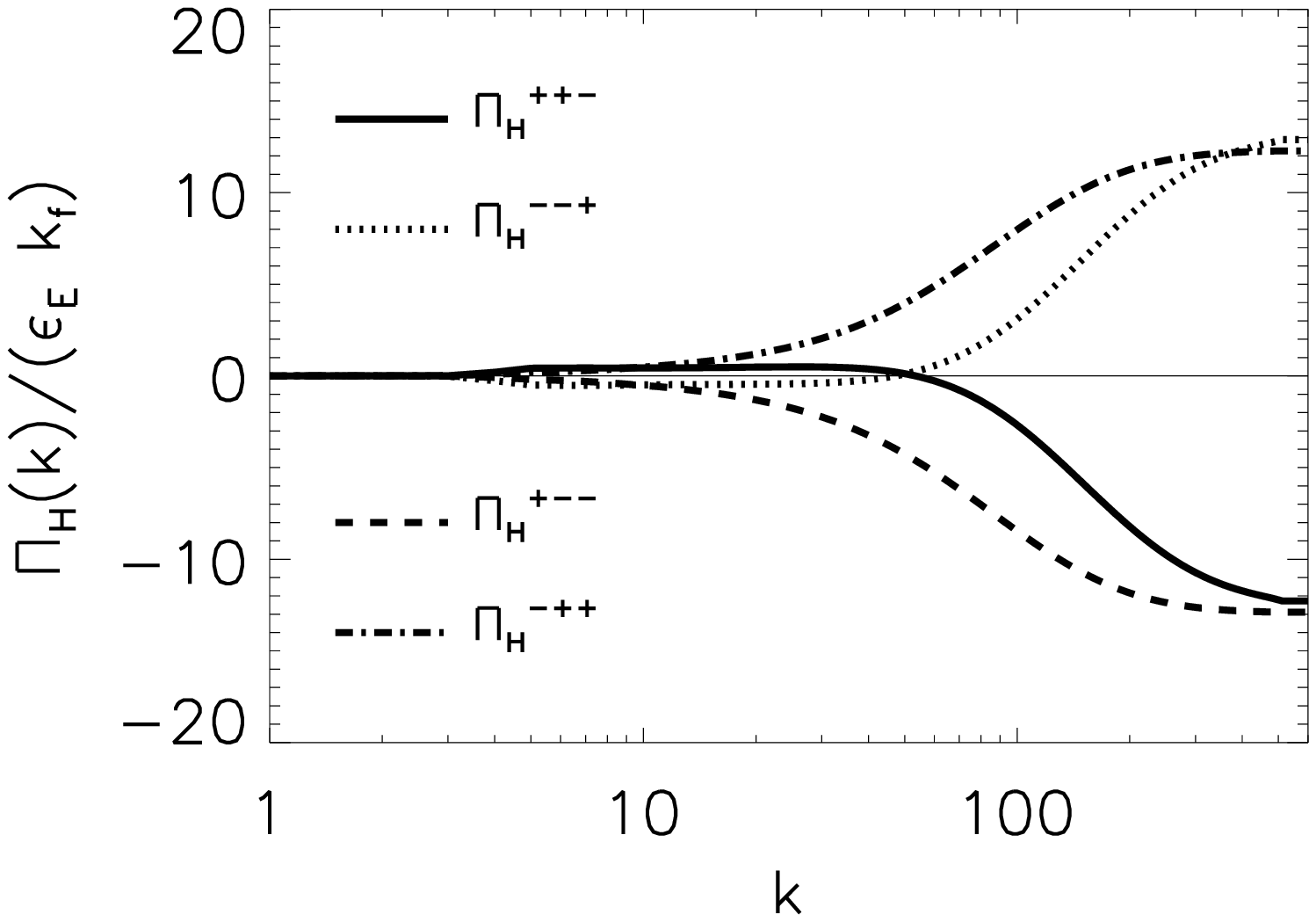}
  \includegraphics[width=7cm]{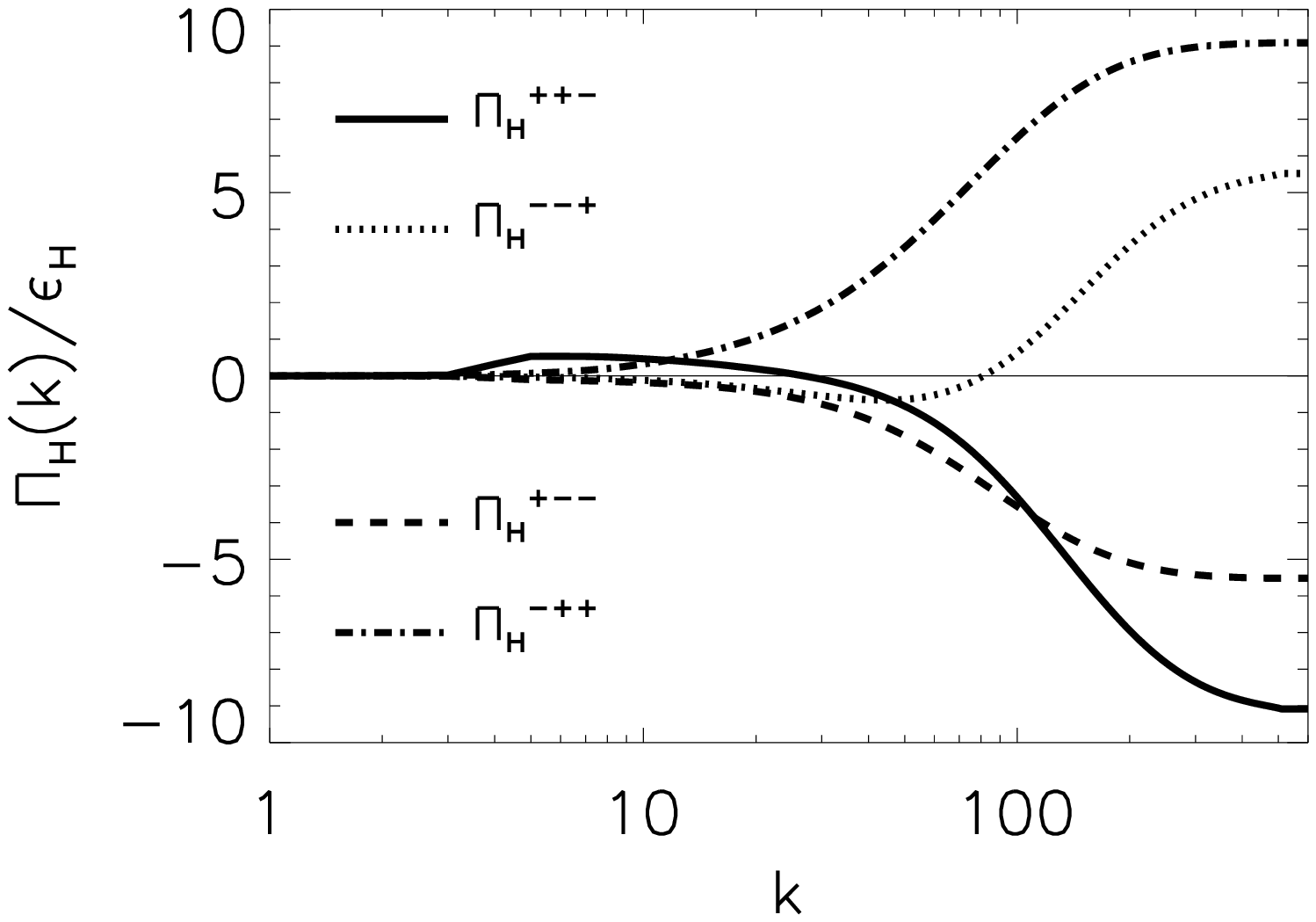}}
  \caption{The four trans-helical helicity fluxes 
  for the non-helical (left) and the helical (right) case:
  $\Pi_{_H}^{++-}$ (solid  line), 
  $\Pi_{_H}^{--+}$ (doted  line),
  $\Pi_{_H}^{+--}$ (dashed line),
  $\Pi_{_H}^{-++}$ (dash dot line).  }
\label{fig:flux_5}
\end{figure}

The trans-helical helicity fluxes for the non-helical (left) and the helical (right) flow are shown in figure \ref{fig:flux_5}.
The $\Pi_H^{++-}$ flux is positive at the inertial range and this implies that these interactions decrease $H^+$.  
On the contrary $\Pi_H^{+--}$ is negative in the inertial range implying an increase of $H^+$ at this range. 
For the non-helical flow these effects balance each other while for the helical case there is a dominance of the fluxes 
that involve interactions with more $u^+$ modes. The same conclusions can be drawn for $H^-$ and the fluxes $\Pi_H^{--+}$ and $\Pi_H^{-++}$. 
In the limit $k\to \infty$ both $\Pi_H^{++-}$ and $\Pi_H^{+--}$ are negative implying net generation of $H^+$ and the terms 
$\Pi_H^{--+}$ and $\Pi_H^{-++}$ are positive implying net generation of $H^-$. 
Thus while there is no net generation of helicity 
there is a net generation of $H^+$ and $H^-$ given by:
\beq \mathcal{G}_{_H} = \lim_{k\to\infty} (\Pi_H^{-++}+\Pi_H^{--+}) = -\lim_{k\to\infty} (\Pi_H^{++-}+\Pi_H^{+--})  >0 .\eeq
This is true both for the helical and the non-helical case. This large generation of simultaneous $H^+$ and $H^-$ 
causes the fluxes  shown in figure \ref{fig:flux_4} to be much larger than the injection rates.

\begin{figure}
  \centerline{
  \includegraphics[width=7cm]{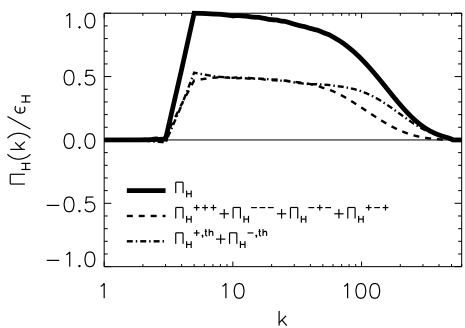} }
  \caption{ Total helicity flux $\Pi_{_H}$ decomposed to $\Pi_{_H}^{+++}+  \Pi_{_H}^{---}+  \Pi_{_H}^{+-+}+\Pi_{_H}^{-+-}$ (dashed line)
  and $\Pi_{_H}^{+,th}+\Pi_{_H}^{-,th} $ (dashed-dot line). }
\label{fig:flux_6}
\end{figure}

Despite, the simultaneous generation of $H^+$ and $H^-$ it turns out that the total helicity flux can also be decomposed 
to two fluxes that remain constant in the inertial range. This is demonstrated in figure \ref{fig:flux_6} where the total helicity flux
is plotted for the helical flow along with the symmetrized fluxes  $\Pi_{_H}^{+++}+  \Pi_{_H}^{---}+  \Pi_{_H}^{+-+}+\Pi_{_H}^{-+-}$
and $\Pi_{_H}^{+,th}+\Pi_{_H}^{-,th} $. Note that for the helicity flux we need to add all non-trans-helical fluxes to obtain 
constant flux in the inertial range. Considering just  $\Pi_{_H}^{+++}+  \Pi_{_H}^{---}$ or just $\Pi_{_H}^{+-+}+\Pi_{_H}^{-+-}$ does not lead
to a constant flux. We also note that adding these fluxes for the non-helical flow leads to zero flux, so it is not displayed.
Again, this is not a trivial result. Conservation of helicity implies only that the total helicity flux is constant 
and we could not {\it a priori} have concluded this result. 

Implications of this result is discussed in the next section where the conclusions are drawn.

\section{Conclusions}                                               

 
In this work we investigated hydrodynamic turbulence using the helical Fourier mode bases proposed in \cite{Lesieur72,Constantin1988}. 
Using this base a decomposition of the energy and helicity fluxes in a turbulent hydrodynamic flow was derived that allowed to investigate separately 
the role of interactions among modes of different helicity in a fully turbulent flow.
This allowed in part to test the predictions of \cite{waleffe1992nature} and the ``instability assumption'' used in that work. 
In the present formalism eight partial energy   fluxes and
                         eight partial helicity fluxes were defined that measure the rate nonlinear interactions
of the particular type transfer energy and helicity from a given spherical set of Fourier modes.
%

The proposed formalism was then applied to the results of large resolution numerical simulations.
Two flows were considered, one without mean helicity and one that was positively helical.
For these flows the partial fluxes were explicitly calculated at steady state.
 The      results are very intriguing.
      As shown in figure \ref{fig:flux_1} the partial energy fluxes defined can be grouped together so that the total flux can be decomposed in three 
      fluxes that are independently constant in the inertial range. This is a nontrivial result as it can not be derived from energy conservation alone
      that implies constancy of only the total energy flux.
      Furthermore, the relative amplitude of these fluxes was the same for both the helical and the non-helical flow 
      and thus these fractions are possibly universal.
      In particular, one of these fluxes that corresponds to same helicity interactions is negative at all scales implying the presence of an inverse cascade of energy,
      that coexists but is overwhelmed by the forward cascade. 
      The helicity flux,  shown in figure \ref{fig:flux_6}, can also be decomposed into two fluxes that are independently constant in the inertial range,
      and are both positive.

The present results have various implications for future investigations
both practical but also theoretical.
First of all, the present results indicate that some of the assumptions for small scale turbulence modeling that should be reviewed.
The presence of an `hidden' inverse cascade (expressed by the negative fluxes of energy $\Pi_E^{+++}$ and $\Pi_E^{---}$)
implies that there is information from the small scales that travels back to the small large scales. The traditional point of view
of small scale modeling assumes that the small scale turbulent motions act only as a sink of turbulent energy transferring it to even 
smaller scales, and thus they are typically modeled as an eddy dissipation term. 

Second,  this investigation indicates how the scales larger than the forcing scale reach an equilibrium. 
The 
injection energy from the small scales driven by the $\Pi^{s,s,s}$ fluxes is balanced by the removal of energy 
from the remaining fluxes. This process might shed light in the deviations observed in the large scale energy spectrum
\citep{dallas2015statistical} from the isothermal equilibrium proposed in \citep{Kraichnan1973}.

%


Finally, we would like to enrich the set of numerical experiments proposed in \citep{Biferale2013jfm} of modified versions of the Navier-Stokes 
as in eq. \ref{eq:bif} by considering the following generalized Navier-Stokes equation: 
\begin{equation}
\partial_t {\bf u}^{s_1} =  \sum_{s_2,s_3}
                            \alpha^{s_1,s_2,s_3}
                            \mathbb{P}^{s_1} \left[ {\bf u}^{s_2} \times {\bf w}^{s_3} \right] 
                         +  \nu \Delta  {\bf u}^{s_1} +  \mathbb{P}^{s_1}[{\bf F}]
                         \label{ANSHC}
\end{equation}
where $\alpha^{s_1,s_2,s_3}$ is a real $2\times2\times2$ matrix. One can then consider a continuous variation from the 
Navier-Stokes obtained for $\alpha^{s_1,s_2,s_3}=1$ to different possible limits. For example one can consider the case where
   the two energies   $E^\pm$ are conserved independently but not the helicity (for $\alpha^{s,-s,s_3}=0$ and the remaining values of $\alpha^{s_1,s_2,s_3}$ are unity)
or the two helicities $H^\pm$ are conserved but not the total energy (for $\alpha^{s,s_2,-s}=0$ and the remaining values of $\alpha^{s_1,s_2,s_3}$ are unity).
Of particular interest is the case for which
$\alpha^{s_1,s_2,s_3}=\lambda$ for all values $s_i$ except when $s_1=s_2=s_3$ for which  $\alpha^{s,s,s}=1$.
Then $\lambda=0$ reduces the system \ref{ANSHC} to \ref{eq:bif}. One could thus continuously transition varying $\lambda$ from a system that cascades
energy forward to a system that cascades energy inversely. Such systems are known to exhibit critical transitions 
\citep{Celani2010,Deusebio2014,Seshasayanan2014,sozza2015dimensional,Seshasayanan2016}. If this is the case it can open new venues for exploring the Navier-Stokes turbulence
as an out-of equilibrium system close to criticality.
%

\acknowledgments

This work was granted access to the HPC resources of MesoPSL financed by the Region Ile de France and the project Equip@Meso (reference ANR-10-EQPX-29-01) of
the programme Investissements d'Avenir supervised by the Agence Nationale pour la Recherche and the HPC resources of GENCI-TGCC-CURIE \& GENCI-CINES-OCCIGEN
(Project No. x2015056421 \& No. x2016056421) where the present numerical simulations have been performed.


\bibliographystyle{jfm}
\bibliography{manuscript_HDT_v01}

\begin{thebibliography}{33}
\expandafter\ifx\csname natexlab\endcsname\relax\def\natexlab#1{#1}\fi
\def\au#1{#1} \def\ed#1{#1} \def\yr#1{#1}\def\at#1{#1}\def\jt#1{\textit{#1}}
  \def\bt#1{#1}\def\bvol#1{\textbf{#1}} \def\vol#1{#1} \def\pg#1{#1}
  \def\publ#1{#1}\def\arxiv#1{#1}\def\org#1{#1}\def\st#1{\textit{#1}}

\bibitem[{Alexakis} {\em et~al.\/}(2005){Alexakis}, {Mininni} \&
  {Pouquet}]{Alexakis2005}
{\sc \au{{Alexakis}, A.}, \au{{Mininni}, P.~D.} \& \au{{Pouquet}, A.}}
  \yr{2005}  \at{{Imprint of Large-Scale Flows on Turbulence}}.  \jt{Physical
  Review Letters}  \bvol{95}~(26),  \pg{264503}.

\bibitem[{Biferale} {\em et~al.\/}(2012){Biferale}, {Musacchio} \&
  {Toschi}]{Biferale2012prl}
{\sc \au{{Biferale}, L.}, \au{{Musacchio}, S.} \& \au{{Toschi}, F.}} \yr{2012}
  \at{{Inverse Energy Cascade in Three-Dimensional Isotropic Turbulence}}.
  \jt{Physical Review Letters}  \bvol{108}~(16),  \pg{164501}.

\bibitem[{Biferale} {\em et~al.\/}(2013){Biferale}, {Musacchio} \&
  {Toschi}]{Biferale2013jfm}
{\sc \au{{Biferale}, L.}, \au{{Musacchio}, S.} \& \au{{Toschi}, F.}} \yr{2013}
  \at{{Split energy-helicity cascades in three-dimensional homogeneous and
  isotropic turbulence}}.  \jt{Journal of Fluid Mechanics}  \bvol{730},
  \pg{309--327}.

\bibitem[{Brandenburg} \& {Subramanian}(2005)]{Brandenburg2005}
{\sc \au{{Brandenburg}, A.} \& \au{{Subramanian}, K.}} \yr{2005}
  \at{{Astrophysical magnetic fields and nonlinear dynamo theory}}.  \jt{Phys.
  Reports}  \bvol{417},  \pg{1--209}.

\bibitem[Cambon \& Jacquin(1989)]{cambon1989spectral}
{\sc \au{Cambon, C} \& \au{Jacquin, L}} \yr{1989}  \at{Spectral approach to
  non-isotropic turbulence subjected to rotation}.  \jt{Journal of Fluid
  Mechanics}  \bvol{202},  \pg{295--317}.

\bibitem[Celani {\em et~al.\/}(2010)Celani, Musacchio \& Vincenzi]{Celani2010}
{\sc \au{Celani, Antonio}, \au{Musacchio, Stefano} \& \au{Vincenzi, Dario}}
  \yr{2010}  \at{Turbulence in more than two and less than three dimensions}.
  \jt{Phys. Rev. Lett.}  \bvol{104},  \pg{184506}.

\bibitem[{Constantin} \& {Majda}(1988)]{Constantin1988}
{\sc \au{{Constantin}, P.} \& \au{{Majda}, A.}} \yr{1988}  \at{{The Beltrami
  spectrum for incompressible fluid flows}}.  \jt{Communications in
  Mathematical Physics}  \bvol{115},  \pg{435--456}.

\bibitem[Dallas {\em et~al.\/}(2015)Dallas, Fauve \&
  Alexakis]{dallas2015statistical}
{\sc \au{Dallas, Vassilios}, \au{Fauve, Stephan} \& \au{Alexakis, Alexandros}}
  \yr{2015}  \at{Statistical equilibria of large scales in dissipative
  hydrodynamic turbulence}.  \jt{Physical Review Letters}  \bvol{115}~(20),
  \pg{204501}.

\bibitem[Deusebio {\em et~al.\/}(2014)Deusebio, Boffetta, Lindborg \&
  Musacchio]{Deusebio2014}
{\sc \au{Deusebio, E.}, \au{Boffetta, G.}, \au{Lindborg, E.} \& \au{Musacchio,
  S.}} \yr{2014}  \at{Dimensional transition in rotating turbulence}.
  \jt{Phys. Rev. E}  \bvol{90},  \pg{023005}.

\bibitem[Falkovich(1994)]{falkovich1994bottleneck}
{\sc \au{Falkovich, Gregory}} \yr{1994}  \at{Bottleneck phenomenon in developed
  turbulence}.  \jt{Physics of Fluids}  \bvol{6}~(4),  \pg{1411}.

\bibitem[Fj{\o}rtoft(1953)]{fjortoft1953changes}
{\sc \au{Fj{\o}rtoft, Ragnar}} \yr{1953}  \at{On the changes in the spectral
  distribution of kinetic energy for twodimensional, nondivergent flow}.
  \jt{Tellus}  \bvol{5}~(3),  \pg{225--230}.

\bibitem[Frisch(1995)]{frisch1995turbulence}
{\sc \au{Frisch, Uriel}} \yr{1995} {\em Turbulence: the legacy of AN
  Kolmogorov\/}.  \publ{Cambridge university press}.

\bibitem[Gilbert(2002)]{gilbert2002magnetic}
{\sc \au{Gilbert, Andrew}} \yr{2002}  \at{Magnetic helicity in fast dynamos}.
  \jt{Geophysical \& Astrophysical Fluid Dynamics}  \bvol{96}~(2),
  \pg{135--151}.

\bibitem[Gilbert \& Childress(1995)]{gilbert1995stretch}
{\sc \au{Gilbert, AD} \& \au{Childress, S}} \yr{1995} Stretch, twist, fold, the
  fast dynamo.

\bibitem[Herring {\em et~al.\/}(1982)Herring, Schertzer, Lesieur, Newman,
  Chollet \& Larcheveque]{herring1982comparative}
{\sc \au{Herring, JR}, \au{Schertzer, D}, \au{Lesieur, M}, \au{Newman, GR},
  \au{Chollet, JP} \& \au{Larcheveque, M}} \yr{1982}  \at{A comparative
  assessment of spectral closures as applied to passive scalar diffusion}.
  \jt{Journal of Fluid Mechanics}  \bvol{124},  \pg{411--437}.

\bibitem[{Kraichnan}(1973)]{Kraichnan1973}
{\sc \au{{Kraichnan}, R.~H.}} \yr{1973}  \at{{Helical turbulence and absolute
  equilibrium}}.  \jt{Journal of Fluid Mechanics}  \bvol{59},  \pg{745--752}.

\bibitem[Kurien {\em et~al.\/}(2004)Kurien, Taylor \&
  Matsumoto]{kurien2004cascade}
{\sc \au{Kurien, Susan}, \au{Taylor, Mark~A} \& \au{Matsumoto, Takeshi}}
  \yr{2004}  \at{Cascade time scales for energy and helicity in homogeneous
  isotropic turbulence}.  \jt{Physical Review E}  \bvol{69}~(6),  \pg{066313}.

\bibitem[Lesieur(1972)]{Lesieur72}
{\sc \au{Lesieur, M.}} \yr{1972}  \bt{D{\'e}composition d’un champ de vitesse
  non divergent en ondes d'h{\'e}licit{\'e}}. {\em Tech. Rep.\/}.
  \org{Observatoire de Nice}.

\bibitem[Lohse \& M{\"u}ller-Groeling(1995)]{lohse1995bottleneck}
{\sc \au{Lohse, Detlef} \& \au{M{\"u}ller-Groeling, Axel}} \yr{1995}
  \at{Bottleneck effects in turbulence: scaling phenomena in r versus p space}.
   \jt{Physical review letters}  \bvol{74}~(10),  \pg{1747}.

\bibitem[Martinez {\em et~al.\/}(1997)Martinez, Chen, Doolen, Kraichnan, Wang
  \& Zhou]{martinez1997energy}
{\sc \au{Martinez, DO}, \au{Chen, S}, \au{Doolen, GD}, \au{Kraichnan, RH},
  \au{Wang, L-P} \& \au{Zhou, Y}} \yr{1997}  \at{Energy spectrum in the
  dissipation range of fluid turbulence}.  \jt{Journal of plasma physics}
  \bvol{57}~(01),  \pg{195--201}.

\bibitem[Mininni {\em et~al.\/}(2006)Mininni, Alexakis \&
  Pouquet]{Mininni2006large}
{\sc \au{Mininni, P.~D.}, \au{Alexakis, A.} \& \au{Pouquet, A.}} \yr{2006}
  \at{Large-scale flow effects, energy transfer, and self-similarity on
  turbulence}.  \jt{Phys. Rev. E}  \bvol{74},  \pg{016303}.

\bibitem[Mininni {\em et~al.\/}(2011)Mininni, Rosenberg, Reddy \&
  Pouquet]{mininni2011hybrid}
{\sc \au{Mininni, Pablo~D}, \au{Rosenberg, Duane}, \au{Reddy, Raghu} \&
  \au{Pouquet, Annick}} \yr{2011}  \at{A hybrid mpi--openmp scheme for scalable
  parallel pseudospectral computations for fluid turbulence}.  \jt{Parallel
  Computing}  \bvol{37}~(6),  \pg{316--326}.

\bibitem[Moffatt(1969)]{moffatt1969degree}
{\sc \au{Moffatt, Henry~Keith}} \yr{1969}  \at{The degree of knottedness of
  tangled vortex lines}.  \jt{J. Fluid Mech}  \bvol{35}~(1),  \pg{117--129}.

\bibitem[{Moffatt}(2014)]{Moffatt2014note}
{\sc \au{{Moffatt}, H.~K.}} \yr{2014}  \at{{Note on the triad interactions of
  homogeneous turbulence}}.  \jt{Journal of Fluid Mechanics}  \bvol{741},
  \pg{R3}.

\bibitem[Sahoo \& Biferale(2015)]{sahoo2015disentangling}
{\sc \au{Sahoo, Ganapati} \& \au{Biferale, Luca}} \yr{2015}  \at{Disentangling
  the triadic interactions in navier-stokes equations}.  \jt{The European
  Physical Journal E}  \bvol{38}~(10),  \pg{1--8}.

\bibitem[Sahoo {\em et~al.\/}(2015)Sahoo, Bonaccorso \&
  Biferale]{sahoo2015role}
{\sc \au{Sahoo, Ganapati}, \au{Bonaccorso, Fabio} \& \au{Biferale, Luca}}
  \yr{2015}  \at{Role of helicity for large-and small-scale turbulent
  fluctuations}.  \jt{Physical Review E}  \bvol{92}~(5),  \pg{051002}.

\bibitem[Seshasayanan \& Alexakis(2016)]{Seshasayanan2016}
{\sc \au{Seshasayanan, Kannabiran} \& \au{Alexakis, Alexandros}} \yr{2016}
  \at{Critical behavior in the inverse to forward energy transition in
  two-dimensional magnetohydrodynamic flow}.  \jt{Phys. Rev. E}  \bvol{93},
  \pg{013104}.

\bibitem[Seshasayanan {\em et~al.\/}(2014)Seshasayanan, Benavides \&
  Alexakis]{Seshasayanan2014}
{\sc \au{Seshasayanan, Kannabiran}, \au{Benavides, Santiago~Jose} \&
  \au{Alexakis, Alexandros}} \yr{2014}  \at{On the edge of an inverse cascade}.
   \jt{Phys. Rev. E}  \bvol{90},  \pg{051003}.

\bibitem[Sozza {\em et~al.\/}(2015)Sozza, Boffetta, Muratore-Ginanneschi \&
  Musacchio]{sozza2015dimensional}
{\sc \au{Sozza, A}, \au{Boffetta, G}, \au{Muratore-Ginanneschi, P} \&
  \au{Musacchio, Stefano}} \yr{2015}  \at{Dimensional transition of energy
  cascades in stably stratified forced thin fluid layers}.  \jt{Physics of
  Fluids (1994-present)}  \bvol{27}~(3),  \pg{035112}.

\bibitem[{Stepanov} {\em et~al.\/}(2015){Stepanov}, {Golbraikh}, {Frick} \&
  {Shestakov}]{stepanov2015hindered}
{\sc \au{{Stepanov}, R.}, \au{{Golbraikh}, E.}, \au{{Frick}, P.} \&
  \au{{Shestakov}, A.}} \yr{2015}  \at{{Hindered Energy Cascade in Highly
  Helical Isotropic Turbulence}}.  \jt{Physical Review Letters}
  \bvol{115}~(23),  \pg{234501}.

\bibitem[{Vainshtein} \& {Zel'dovich}(1972)]{Vainshtein1972}
{\sc \au{{Vainshtein}, S.~I.} \& \au{{Zel'dovich}, Y.~B.}} \yr{1972}  \at{{On
  the origin of magnetic fields in astrophysics. (Turbulence mechanisms
  ''dynamo'').}}  \jt{Uspekhi Fizicheskikh Nauk}  \bvol{106},  \pg{431--457}.

\bibitem[Verma {\em et~al.\/}(2005)Verma, Ayyer, Debliquy, Kumar \&
  Chandra]{verma2005local}
{\sc \au{Verma, Mahendra~K}, \au{Ayyer, Arvind}, \au{Debliquy, Olivier},
  \au{Kumar, Shishir} \& \au{Chandra, Amar~V}} \yr{2005}  \at{Local
  shell-to-shell energy transfer via nonlocal interactions in fluid
  turbulence}.  \jt{Pramana}  \bvol{65}~(2),  \pg{297--310}.

\bibitem[{Waleffe}(1992)]{waleffe1992nature}
{\sc \au{{Waleffe}, F.}} \yr{1992}  \at{{The nature of triad interactions in
  homogeneous turbulence}}.  \jt{Physics of Fluids}  \bvol{4},  \pg{350--363}.

\end{thebibliography}

\end{document}